\documentclass[11pt]{article}

\usepackage[a4paper,margin=2.5cm]{geometry}
\usepackage{graphicx}
\usepackage{amsmath}
\usepackage{amssymb}
\usepackage{bm}
\usepackage{dcolumn}
\usepackage{ragged2e}
\usepackage{microtype}
\usepackage{etoolbox}
\usepackage{xcolor}
\usepackage{harvard}
\usepackage{hyperref}

\hypersetup{
    colorlinks=true,
    citecolor=blue,
    linkcolor=blue,
    urlcolor=blue
}

\makeatletter
\renewcommand{\harvardcite}[4]{%
  \global\@namedef{HAR@fn@#1}{#2}%
  \global\@namedef{HAR@an@#1}{#3}%
  \global\@namedef{HAR@yr@#1}{\hyper@@link[cite]{}{cite.#1}{#4}}%
  \global\@namedef{HAR@df@#1}{\csname HAR@fn@#1\endcsname}%
}
\makeatother

\definecolor{revisionblue}{RGB}{0,0,0}

\AtBeginEnvironment{abstract}{\justifying}

\begin{document}

\begin{center}

{\LARGE\bfseries
Hamiltonian dynamics for stochastic reconstruction in emission tomography
\par}

\vspace{1em}

T Leontiou$^{1,2,*}$,
A Frixou$^{3}$,
E Ttofi$^{4,5}$,
C Chrysostomou$^{6}$,
Y Parpottas$^{1,2}$,
K Michael$^{7}$,
S Frangos$^{7}$,
E Stiliaris$^{3,8}$ and
C N Papanicolas$^{3,9}$

\vspace{1em}

\small
$^1$Department of Mechanical Engineering, Frederick University, Nicosia, Cyprus\\
$^2$Frederick Research Center, Nicosia, Cyprus\\
$^3$CaSToRC, The Cyprus Institute, Nicosia, Cyprus\\
$^4$Department of Electrical and Computer Engineering and Informatics, Frederick University, Nicosia, Cyprus\\
$^5$Information Technology Department, German Medical Institute, Limassol, Cyprus\\
$^6$ERATOSTHENES Centre of Excellence, Cyprus University of Technology, Limassol, Cyprus\\
$^7$Bank of Cyprus Oncology Center, Nicosia 2006, Cyprus\\
$^8$Department of Physics, National and Kapodistrian University of Athens, Athens, Greece\\
$^9$The Cyprus Academy of Sciences, Letters and Arts\\[0.5em]
$^*$Corresponding author: t.leontiou@frederick.ac.cy

\end{center}

\vspace{1em}
% ===== BEGIN SUBSTANTIVE REVISION =====
\begin{abstract}
\justifying
\textit{Objective.}
To develop a practical stochastic reconstruction framework for emission
tomography that generates ensembles of data-compatible images and enables
uncertainty quantification and assessment of forward-model adequacy. \textit{Approach.}
The framework combines stochastic-gradient descent initialization with
Hamiltonian Monte Carlo (HMC) sampling directly in high-dimensional voxel
space. Beyond point reconstruction, we introduce a spatially resolved
operator-weighted diagnostic, the sampled data-visible variance, which
quantifies how image fluctuations propagate through the imaging operator
and thereby probes the local conditioning of the inverse problem under
realistic acquisition physics. The methodology is evaluated using
controlled software phantoms, experimental anthropomorphic phantom
measurements, and a clinical DATSCAN SPECT acquisition. \textit{Main results.}
Under ideal conditions, the HMC ensemble mean provides point-estimate
accuracy comparable to deterministic reconstruction methods, while the
sampled ensemble provides additional physically interpretable information.
The ensemble analysis helps distinguish uncertainty associated with the
intrinsic ill-posedness of the inverse problem from variability linked to
forward-model inadequacy. The clinical example demonstrates applicability
under realistic acquisition statistics rather than diagnostic performance.
\textit{Significance.}
The proposed stochastic reconstruction framework provides a practical
ensemble-based approach for emission tomography that extends conventional
point reconstruction with model-conditioned uncertainty estimates and
spatially resolved diagnostics of forward-model adequacy.

\end{abstract}

\section{Introduction}

Emission tomography poses a computationally intensive inverse problem: reconstructing the
spatial distribution of radiotracer activity from projection data that are incomplete,
noisy, and affected by physical processes including photon attenuation and scattering.
To place this challenge in context, the imaging process can be summarized in a series of
key steps: % (Figure~\ref{fig:block_diagram}):
1. a radiotracer is injected into the patient, where it undergoes uptake by tissues and
   radioactive decay with $\gamma$-ray emission;
2. the emitted photons propagate through the body, undergoing attenuation and scattering;
3. photons are detected by a collimated gamma camera, which restricts their directionality;
4. the detected counts are organized into projection data, known as sinograms;
5. finally, the inverse problem is solved to reconstruct the underlying radiotracer
   distribution from the measured sinograms.

% \begin{figure*}[th]
% \centering
% \includegraphics[width=0.9\textwidth]{figures/block_diagram.png}
% \caption{Key steps of the emission tomography process: from radiotracer administration and
% photon propagation, through detection and projection data formation, to the final inverse
% reconstruction problem.}
% \label{fig:block_diagram}
% \end{figure*}

% ===== BEGIN SUBSTANTIVE REVISION =====
Traditionally, image reconstruction in emission tomography has relied on
deterministic iterative methods such as Maximum Likelihood Expectation
Maximization (MLEM), Ordered Subsets Expectation Maximization (OSEM), and
penalized or regularized variants~\cite{diagnostics14131431}. These
methods produce a single image estimate by optimizing an objective
function derived from the statistical model of the measured projection
data.
% {
% \color{revisionblue}
Although such approaches are well established and clinically
successful, they do not by themselves provide an ensemble of admissible
images or a direct characterization of the variability associated with
the reconstruction.

Several stochastic and Bayesian approaches have been developed to address
this limitation from different perspectives. 
% Sitek~\cite{sitek2011origin}
\citeasnoun{sitek2011origin}
introduced the origin-ensemble approach for emission tomography, in which
the detected events are associated with possible locations of origin and
an ensemble average over these origin assignments is used to estimate the
activity distribution. This formulation provides a statistical
reconstruction method with a different computational structure from
standard likelihood-based iterative algorithms. The work shows that, under
conditions relevant to practical emission tomography, the origin-ensemble
estimate closely approximates the maximum-likelihood estimate, while
offering a useful alternative view of reconstruction in terms of event
origins.

A second line of work has focused on rigorous Bayesian posterior
formulations for emission tomography. 
% Zhou \emph{et al.}~\cite{zhou2020bayesian}
\citeasnoun{zhou2020bayesian}
developed an infinite-dimensional Bayesian framework for image
reconstruction with Poisson data. Their work combines a Poisson data
model with a regularizing prior and develops a sampling strategy designed
to remain stable under discretization refinement. The emphasis of this
work is mathematical and statistical: the authors establish a well-posed
Bayesian formulation, define a posterior distribution over images, and
use samples from this posterior to quantify image uncertainty and detect
possible artifacts.

Filipovi'{c} \emph{et al.}~\cite{filipovic2019posterior,filipovic2021posterior}
% \citeasnoun{filipovic2019posterior,filipovic2021posterior}
studied posterior image distributions in PET using Gibbs sampling and,
more recently, the posterior bootstrap. In the posterior-bootstrap
approach, approximate posterior image realizations are generated by
randomly perturbing the data likelihood and repeatedly solving the
corresponding reconstruction problem. This provides a scalable and
parallelizable way to obtain posterior means, variances, covariances,
intervals, and ROI-level uncertainty measures for a range of Bayesian PET
models, including models with anatomical information from MRI. A central
point in this work is the distinction between the posterior image
distribution obtained from a single dataset and the variability of
reconstructed estimators over repeated noisy acquisitions.

Together, these studies demonstrate that ensemble and posterior-based
ideas can provide information beyond a single reconstructed image. They
also clarify that different ensembles answer different questions. Origin
ensembles describe uncertainty in the assignment of detected events to
their possible origins. Bayesian posterior methods describe uncertainty
under a specified likelihood, prior, and reconstruction model.
Posterior-bootstrap methods provide an efficient approximation to image
posterior distributions by repeatedly solving randomized reconstruction
problems. These approaches are therefore primarily concerned with the
construction, sampling, approximation, or interpretation of image
distributions under a chosen statistical model.

In parallel, recent years have witnessed rapid progress in machine learning--based
reconstruction and image synthesis methods for emission tomography~\cite{wang2020dlrecon,apostolopoulos2023spectdl}.
These approaches have demonstrated substantial gains in
computational speed and image quality through data-driven training and hardware-optimized
implementations, and are increasingly attractive for clinical deployment.
In addition to direct reconstruction methods, a growing body of work has explored
cross-modality image generation, where PET images are synthesized from anatomical
modalities such as MRI using deep generative models.
In particular, diffusion-based approaches have recently shown strong performance in
conditional image generation tasks.
% Cong \emph{et al.}~\cite{cong2025diffusion}
\citeasnoun{cong2025diffusion}
proposed a multi-sequence, multi-tracer framework in which PET images are generated from
multiple MRI sequences using cross-attention mechanisms and tracer-specific embeddings,
enabling a single model to capture correlations across imaging modalities and tracer types.

% { \color{revisionblue}
Against this background, the present work builds on the Athens Model
Independent Analysis Scheme (AMIAS)~\cite{amias_baryons,amias_papan} and
its tomographic extension, the Reconstructed Image from Simulations
Ensemble (RISE) method~\cite{papanicolas2018novel}. 
AMIAS formulates
inverse problems in terms of ensembles of admissible solutions whose
probabilities are determined by their agreement with measured data. The
method was originally developed in hadronic physics
\cite{amias_LQCD1,amias_LQCD2,amias_markou} and was later adapted to
medical imaging~\cite{loizos_phd}. In RISE, the voxel intensities of a
reconstructed image are treated as the parameters of a high-dimensional
probability density, and complete images are sampled according to their
consistency with the measured projections. Previous RISE applications in
SPECT~\cite{koutsantonis2018rise,keliri2021myocardial},
PET~\cite{lemesios2021pet}, and infrared emission
tomography~\cite{koutsantonis2019infrared} demonstrated the conceptual
value of ensemble-based reconstruction and uncertainty estimation
\cite{papanicolas2020uncertainties}.

The present work reformulates this AMIAS/RISE philosophy using modern
stochastic optimization and Hamiltonian Monte Carlo (HMC) sampling in
finite voxel space. The aim is not to construct a fully rigorous
Bayesian posterior in the sense of the Bayesian works discussed above,
nor to approximate a posterior distribution through repeated randomized
deterministic reconstructions. Instead, the objective is to generate an ensemble of
data-compatible reconstructed images under a specified SPECT forward
model, and to use this ensemble to study how reconstruction fluctuations
interact with the measurement physics.

HMC is used because the adopted likelihood-based sampling density is
differentiable and can be explored efficiently using gradient information.
This enables long-range proposals in the high-dimensional voxel space and
reduces the local random-walk behavior of simpler samplers. Stochastic
gradient descent first locates a high-probability region, after which HMC
generates an ensemble of admissible images around it.

The main distinction of the present work is the use of this ensemble for
forward-model sensitivity analysis. Conventional posterior, bootstrap, or
image-space ensemble variance quantifies how much reconstructed images vary
under a chosen sampling procedure. The diagnostics developed here instead
test whether those fluctuations remain visible through the SPECT measurement
operator. The ensemble can therefore quantify image variability and assess
whether forward-model refinements, such as attenuation correction, produce
additional data-coupled changes.

Residual mismatch and ensemble fluctuation probe different properties. A
residual diagnostic tests whether the mean reconstructed image leaves
structured disagreement with the measured data, whereas the data-visible
diagnostic tests whether sampled images fluctuate along directions that
remain observable in projection space. The two are complementary: the first
is mainly sensitive to bias or model inconsistency in the mean image, while
the second measures data-coupled variability.

This framework is not intended to replace deterministic, Bayesian,
posterior-bootstrap, origin-ensemble, or learning-based approaches. It
provides a physics-driven stochastic layer for assessing reconstruction
variability and forward-model adequacy. The emphasis is therefore not on
improving visual appearance, but on determining whether the adopted SPECT
model explains the measured data and whether model refinements produce
measurable changes in the reconstructed ensemble.

The principal contributions of this work are therefore:

\begin{itemize}

\item \textbf{{Efficient voxel-space stochastic reconstruction:}}
A stochastic reformulation that enables practical ensemble generation
for full three-dimensional emission tomography using modern tensor-based
numerical frameworks.

\item \textbf{Application of Hamiltonian Monte Carlo as a reconstruction
engine:}
Reconstruction is performed through simulation of classical Hamiltonian
dynamics, providing a physics-motivated sampling mechanism for
high-dimensional probability distributions.

\item \textbf{Forward-model assessment via ensemble diagnostics:}
The generated image ensembles are used to quantify how fluctuations
propagate through the measurement operator, enabling structured
evaluation of forward-model adequacy beyond global residual metrics.

\item \textbf{Quantitative estimation of observables:}
Once an adequate forward model is specified, physical observables,
including regional activity measures and image-derived ratios, can be
evaluated directly from the generated ensemble, yielding both estimates
and model-conditioned uncertainty estimates.

\end{itemize}

The methodology is demonstrated using controlled software phantoms,
experimental anthropomorphic phantom data, and clinical DATSCAN
acquisitions.

\section{Materials and Methods}

This section presents the methodological framework used for stochastic
reconstruction in emission tomography. We first formulate the inverse
problem and the statistical model used to define the probability
distribution of reconstructed images, while clarifying its relation to
conventional Bayesian posterior formulations and the distinctions relevant
to the present work. The Hamiltonian Monte Carlo (HMC) sampling strategy
used to explore this distribution is then described, together with the
gradient-based optimization procedure employed for initialization.
Finally, we introduce the ensemble-based diagnostics used to analyze
fluctuations and their interaction with the measurement operator.

\subsection{AMIAS probability density for image ensembles}
\label{sec:amias_image_pdf}

Emission tomography is naturally formulated as a statistical inverse
problem because the measured projection data arise from photon-counting
processes. Let \(y_i\) denote the measured number of counts in detector
bin \(i\), and let \(\bm{x}\) denote the unknown activity distribution in
image space. For a linear forward model, the expected number of counts in
bin \(i\) is written as
\begin{equation}
\mu_i(\bm{x})
=
(\bm{P}\bm{x})_i+r_i ,
\label{eq:forward_mean}
\end{equation}
where \(\bm{P}\) is the projection operator used in the reconstruction and
\(r_i\) denotes known additive contributions such as scatter, background,
or other correction terms.
\begingroup
\color{revisionblue}
No separate additive background or scatter sinogram was supplied in any
of the reconstructions reported in this work. Accordingly, all subsequent
reconstruction and analysis expressions are written for the case
\(r_i=0\). Attenuation and, where applicable, scatter effects are nevertheless
present when they are represented by the photon-transport physics of the
corresponding forward model \(\bm{P}\). Scatter or background contributions
present in the measured data but not represented by the adopted forward
model contribute instead to forward-model mismatch.
\endgroup

The standard emission-tomography measurement model assumes independent
Poisson counting statistics,
\begin{equation}
y_i
\sim
\mathrm{Poisson}\!\left(\mu_i(\bm{x})\right),
\end{equation}
with likelihood
\begin{equation}
p(\bm{y}\mid\bm{x})
=
\prod_i
\frac{
\mu_i(\bm{x})^{y_i}
e^{-\mu_i(\bm{x})}
}{y_i!}.
\label{eq:poisson_likelihood_general}
\end{equation}
Conventional maximum-likelihood reconstruction, including MLEM-type
algorithms, is associated with maximizing this likelihood with respect to
the image variables.

For the moderate-to-high count regime considered in the present SPECT
studies, we use a weighted least-squares approximation to the Poisson
likelihood. The corresponding compatibility function is
\begin{equation}
\chi^2(\bm{x})
=
\sum_i
\frac{
\left[
y_i-(\bm{P}\bm{x})_i
\right]^2
}{
\sigma_i^2
},
\label{eq:amias_chi2_diagonal}
\end{equation}
where \(\sigma_i^2\) represents the counting variance in detector bin
\(i\). In the present implementation, the variances are estimated from the
measured counts using a stabilized fixed-weight form,
\begin{equation}
\sigma_i^2
\simeq
y_i+\epsilon ,
\label{eq:poisson_variance_estimate}
\end{equation}
with \(\epsilon>0\) preventing singular weights in empty or near-empty
bins.\footnote{
The same stabilization constant \(\epsilon\) is used throughout the
weighted \(\chi^2\) objective. Its role is numerical rather than
statistical: it prevents divisions by zero in empty or very low-count
bins. For the count levels considered in this work, the results were not
sensitive to the precise value of \(\epsilon\); values between \(10^{-7}\)
and \(1\) gave indistinguishable reconstructed images and ensemble
diagnostics.
}
If detector-bin correlations are important, the diagonal weighting
in Eq.~\eqref{eq:amias_chi2_diagonal} can be replaced by a full covariance
matrix. In the present work, however, the diagonal approximation is used.
\begingroup
\color{revisionblue}
The empirical assessment of the Gaussian weighted least-squares
approximation used in Eq.~\eqref{eq:amias_chi2_diagonal} is reported at
the beginning of Sec.~\ref{sec:empirical_probability_validation}.
\endgroup

The sampling measure over admissible images is then defined as
\begin{equation}
\Pi_{\rm A}(\bm{x}\mid\bm{y})
=
\frac{1}{Z_x}
\mathbf{1}_{\Omega_x}(\bm{x})
\exp\left[
-\frac{1}{2}\chi^2(\bm{x})
\right],
\label{eq:amias_sampling_measure}
\end{equation}
where \(\Omega_x\) denotes the finite numerical domain of admissible image
configurations and
\begin{equation}
Z_x
=
\int_{\Omega_x}
\exp\left[
-\frac{1}{2}\chi^2(\bm{x})
\right]
\,d\bm{x}
\label{eq:amias_normalization}
\end{equation}
is the normalization factor. In the present implementation no explicit
anatomical, smoothing, or structural prior is introduced. The density in
Eq.~\eqref{eq:amias_sampling_measure} should therefore be interpreted as a
likelihood-based compatibility measure over admissible images, rather than
as a fully specified Bayesian posterior with an explicit prior.

For any image observable \(F(\bm{x})\), the expectation value is
\begin{equation}
\langle F\rangle_{\rm A}
=
\frac{
\displaystyle
\int_{\Omega_x}
F(\bm{x})
\exp[-\chi^2(\bm{x})/2]
\,d\bm{x}
}{
\displaystyle
\int_{\Omega_x}
\exp[-\chi^2(\bm{x})/2]
\,d\bm{x}
}.
\label{eq:amias_observable_average}
\end{equation}
In particular, the mean image is
\begin{equation}
\bar{\bm{x}}_{\rm A}
=
\langle \bm{x}\rangle_{\rm A},
\label{eq:amias_mean_image}
\end{equation}
and the voxel-wise ensemble variance is
\begin{equation}
\sigma_{\rm A}^2(x_j)
=
\left\langle
\left(
x_j-\bar{x}_{{\rm A},j}
\right)^2
\right\rangle_{\rm A}.
\label{eq:amias_voxel_variance}
\end{equation}
The mean image should not be identified in general with the
maximum-likelihood image. The latter corresponds to the minimizer of the
data-misfit function, whereas the mean is an expectation value over the
sampled image ensemble.

Regional observables are treated in the same way. For a region \(R\), we
define the normalized regional density
\begin{equation}
\rho_R(\bm{x})
=
\frac{
\displaystyle
\frac{1}{|R|}\sum_{j\in R}x_j
}{
\displaystyle
\frac{1}{|M|}\sum_{j\in M}x_j
},
\label{eq:normalized_regional_density_validation}
\end{equation}
where \(M\) is a fixed normalization mask over the image domain. This
normalization reduces sensitivity to global count-scaling differences and
focuses the comparison on relative reconstructed activity.

The high-dimensional integrals in
Eq.~\eqref{eq:amias_observable_average} are evaluated by Monte Carlo
sampling of Eq.~\eqref{eq:amias_sampling_measure}. In this work,
Hamiltonian Monte Carlo is used to sample the image voxels from the
density defined above. Since the compatibility function is differentiable,
gradient information can be used both to initialize the reconstruction
with a deterministic optimizer and to generate efficient HMC trajectories
in the high-dimensional image space.

\begingroup
\color{revisionblue}
The corresponding empirical validation of the AMIAS sampling scheme,
including repeated-acquisition image-space and projection-space
consistency tests, is presented in
Sec.~\ref{sec:empirical_probability_validation}.
\endgroup

Although the present implementation uses a Gaussian weighted least-squares
compatibility function, the AMIAS/RISE construction is not restricted to
this approximation. Once a forward model and measurement model have been
specified, the sampling density over admissible images can be defined by
the corresponding statistical compatibility factor. In lower-count
regimes, or when a more exact treatment of photon-counting statistics is
required, the Gaussian misfit may be replaced by a Poisson likelihood or
Poisson deviance. Such choices define different image ensembles and
should be validated for the specific acquisition model and statistical
question.

\subsubsection{Hamiltonian Monte Carlo sampling}
\label{sec:hmc_sampling}

The probability density defined in Eq.~\eqref{eq:amias_sampling_measure}
is sampled using Hamiltonian Monte Carlo (HMC). HMC is a
gradient-informed Markov Chain Monte Carlo method that uses concepts from
classical mechanics to explore high-dimensional probability distributions
efficiently~\cite{betancourt2017hmc,HMC1}. Its use here follows naturally
from the differentiable weighted least-squares compatibility function
introduced in the previous section.

The choice of HMC is motivated by the large dimensionality of voxel-based
emission tomography problems, where the number of unknown image
parameters may become very large. In such spaces, local random-walk Monte
Carlo schemes explore configuration space diffusively and can require a
prohibitively large number of likelihood evaluations. HMC instead uses
gradient information and auxiliary momentum variables to generate
long-range proposals while maintaining high acceptance probability. This
is particularly useful in emission tomography, where each sampling
trajectory requires repeated application of the projection and
backprojection operators.

To construct the HMC dynamics, auxiliary momentum variables \(\bm{p}\)
are introduced and the Hamiltonian is written as
\begin{equation}
\mathcal{H}(\bm{x},\bm{p})
=
U(\bm{x})+K(\bm{p}) ,
\label{eq:hmc_hamiltonian}
\end{equation}
where the potential energy is the negative log-density of
Eq.~\eqref{eq:amias_sampling_measure}, up to an additive constant. Inside
the admissible image domain this gives
\begin{equation}
U(\bm{x})
=
\frac{1}{2}\chi^2(\bm{x})
=
\frac{1}{2}
\sum_i
\frac{
\left[
y_i-(\bm{P}\bm{x})_i
\right]^2
}{
\sigma_i^2
},
\label{eq:hmc_potential}
\end{equation}
with the same fixed weights as in Eq.~\eqref{eq:amias_chi2_diagonal}. The
kinetic energy is chosen as
\begin{equation}
K(\bm{p})
=
\frac{1}{2}
\bm{p}^{T}
M^{-1}
\bm{p},
\label{eq:hmc_kinetic}
\end{equation}
where \(M\) is the mass matrix. In the present implementation, \(M\) was
chosen to be diagonal and proportional to the identity, which provided
stable sampling behavior for the systems considered here.

The joint distribution in \((\bm{x},\bm{p})\) is explored by numerically
integrating Hamilton's equations,
\begin{align}
\frac{d\bm{x}}{dt}
&=
\nabla_{\bm{p}}\mathcal{H}
=
M^{-1}\bm{p},
\\
\frac{d\bm{p}}{dt}
&=
-\nabla_{\bm{x}}\mathcal{H}
=
-\nabla_{\bm{x}}U(\bm{x}).
\end{align}
For the weighted least-squares potential in
Eq.~\eqref{eq:hmc_potential}, the gradient is
\begin{equation}
\nabla_{\bm{x}}U(\bm{x})
=
\bm{P}^{T}\bm{W}
\left(
\bm{P}\bm{x}-\bm{y}
\right),
\label{eq:hmc_potential_gradient}
\end{equation}
where
\[
\bm{W}=\mathrm{diag}\left(1/\sigma_i^2\right).
\]
Equivalently, the HMC force term is
\begin{equation}
-\nabla_{\bm{x}}U(\bm{x})
=
\bm{P}^{T}\bm{W}
\left(
\bm{y}-\bm{P}\bm{x}
\right).
\label{eq:hmc_force}
\end{equation}
Thus, each HMC force evaluation requires one forward projection followed
by one weighted backprojection. No numerical finite-difference gradients
are required; the same projection/backprojection structure used in
gradient-based reconstruction is used to evolve the HMC trajectories.
The equations are integrated using the leapfrog integrator.
After a finite number of leapfrog steps, a
Metropolis accept/reject step corrects for numerical integration error and
ensures sampling of the target density.

Since the image parameters represent activity values, the sampling domain
is restricted to non-negative voxel intensities. In the final
implementation this constraint was imposed using reflective boundary
conditions at \(x_j=0\). When a trajectory attempted to cross the boundary
of the admissible domain, the corresponding coordinate and momentum
component were reflected, so that the trajectory remained inside the
non-negative image space. Reflective boundary treatments of this type are
standard in constrained Hamiltonian Monte Carlo and related Hamiltonian
sampling methods~\cite{pakman2014exact}.

As a numerical robustness check, we also tested simpler alternatives in
which proposed negative values were projected back onto the admissible
domain, as well as direct clipping of negative voxel values. Such
projection-based corrections are closely related to constrained
Hamiltonian dynamics and SHAKE/RATTLE-type integrators used for systems
with holonomic constraints~\cite{ryckaert1977numerical,andersen1983rattle,mclachlan2014geometric}.
In the present reconstructions, the choice
between reflection, projection, and clipping did not visibly alter the
reconstructed image quality or the main ensemble-derived statistical
diagnostics. The results reported below use the reflective implementation.

The reconstructed image is obtained as the ensemble expectation value of
the sampled configurations, while variances and other image-derived
observables are computed directly from the same ensemble. In practice, a
deterministic gradient-based optimizer is first used to locate a
high-probability region of the density. The resulting image is then used
to initialize the HMC chains, improving thermalization and reducing the
number of discarded samples.

The use of HMC in high-dimensional statistical systems is well established
in computational physics, including lattice quantum chromodynamics, where
the sampled configuration spaces and correlation structures are highly
non-trivial. In the present tomography application, GPU-accelerated
tensor operations make repeated applications of the projection and
backprojection operators feasible within HMC trajectories, enabling direct
stochastic ensemble generation in voxel space.

The purpose of the HMC ensemble is not limited to producing a mean image.
As discussed in Section~\ref{sec:datavisiblevar}, the sampled
fluctuations can also be propagated through the data-weighted
projection--backprojection operator
\[
\bm{H}
=
\bm{P}^{T}
\bm{W}
\bm{P},
\]
where \(\bm{W}\) contains the statistical weights used in the
data-fidelity term. This operation suppresses fluctuations that are weakly
visible to the measured data and emphasizes the component of the ensemble
variability that remains coupled to the projection measurements. The same
sampled ensemble therefore provides both reconstructed image estimates and
diagnostics of forward-model sensitivity.

Convergence and stability of the HMC chains were assessed using trace
plots, stabilization of ensemble means, acceptance ratios, and consistency
of variance estimates across independent runs. Parallel tempering was also
used to improve thermalization and reduce trapping in local regions of the
high-dimensional density. Additional implementation details, including
leapfrog parameters, acceptance statistics, thermalization, parallel
tempering, and autocorrelation diagnostics, are provided in
Appendix~\ref{app:hmc_implementation}.

\begin{figure*}[th!]
    \centering
    \includegraphics[width=0.95\textwidth]{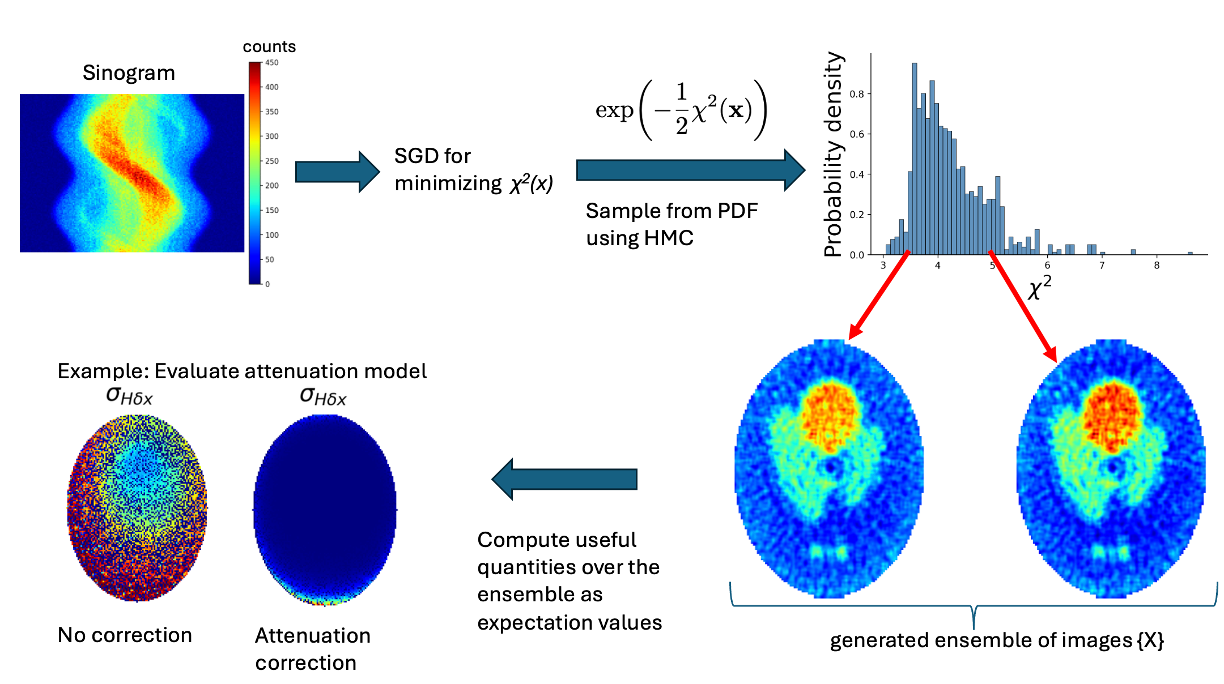}
    \caption{
Overview of the stochastic reconstruction workflow and the concept of
data-visible standard deviation. Starting from the measured detector sinogram
(top left), a likelihood-based density over admissible images is defined
as \(\Pi(\bm{x})\propto\exp[-\chi^2(\bm{x})/2]\) (top middle). A fast
gradient-based reconstruction is first used to locate a high-probability
region of this density, providing initialization for Hamiltonian Monte
Carlo sampling. HMC then generates an ensemble of reconstructed images;
the top-right panel illustrates the resulting \(\chi^2\) distribution,
and the example images correspond to representative samples from the
ensemble. From the generated ensemble, voxel-wise means, variances, and
derived observables can be computed directly. 
The data-visible standard deviation is obtained by propagating ensemble
fluctuations through the data-weighted projection--backprojection operator
\(\bm{H}=\bm{P}^{\top}\bm{W}\bm{P}\), thereby isolating the component of the
sampled variability that remains coupled to the measured projection data.
Comparison of these diagnostics under different forward-model assumptions,
such as with and without attenuation correction, allows the ensemble to be
used for assessing forward-model adequacy beyond global residual measures.
    }
    \label{fig:dvv_overview}
\end{figure*}

\subsubsection{Data-visible variance from image ensembles}
\label{sec:datavisiblevar}

% ===== BEGIN SUBSTANTIVE REVISION =====
% \begingroup
% \color{revisionblue}
Stochastic reconstruction methods such as Hamiltonian Monte Carlo (HMC)
generate an ensemble of images
\(\{\mathbf{x}^{(s)}\}_{s=1}^{N_s}\)
drawn from the sampling distribution
\(\Pi(\mathbf{x})\) defined by
Eq.~\eqref{eq:amias_sampling_measure}. The ensemble contains statistical
information regarding the variability of reconstructed quantities
\cite{papanicolas2020uncertainties,reader2015bayesian}.

In the present analysis, each sampled image is first normalized by its
mean value inside a fixed support mask \(M\),
\begin{equation}
\tilde{\mathbf{x}}^{(s)}
=
\frac{\mathbf{x}^{(s)}}{\langle \mathbf{x}^{(s)}\rangle_M},
\qquad
\langle \mathbf{x}^{(s)}\rangle_M
=
\frac{1}{|M|}
\sum_{j\in M} x_j^{(s)} .
\label{eq:normalized_samples_dvv}
\end{equation}
This sample-wise scalar normalization is used throughout the paper for
all ensemble-derived fluctuation maps and derived observables.
The ensemble mean and sample variance are then
\begin{equation}
\bar{\tilde{\mathbf{x}}}
=
\langle \tilde{\mathbf{x}}\rangle,
\qquad
\sigma^2(\tilde{x}_j)
=
\left\langle
\left(
\tilde{x}_j-\bar{\tilde{x}}_j
\right)^2
\right\rangle ,
\label{eq:image_sample_variance}
\end{equation}
where \(\langle\cdot\rangle\) denotes an average over the sampled images.

To relate image-space fluctuations to their influence on the measured
data, we write the weighted least-squares discrepancy introduced in
Eq.~\eqref{eq:amias_chi2_diagonal} as
\begin{equation}
\chi^2(\mathbf{x})
=
(\mathbf{y}-\mathbf{P}\mathbf{x})^{\top}
\mathbf{W}
(\mathbf{y}-\mathbf{P}\mathbf{x}),
\qquad
\mathbf{W}
=
\operatorname{diag}
\left(
\frac{1}{y_i+\epsilon}
\right).
\label{eq:phi_wls}
\end{equation}

This naturally introduces the operator
\begin{equation}
\bm{H}
=
\bm{P}^{\top}
\bm{W}
\bm{P},
\label{eq:H_def}
\end{equation}
We denote this operator by \(\bm{H}\) for compactness; it is distinct from
the scalar Hamiltonian \(\mathcal{H}(\bm{x},\bm{p})\) used in the HMC
dynamics. The operator \(\bm{H}\) corresponds to the Hessian, or
Gauss--Newton/Fisher approximation of the quadratic data-fidelity term.
Such operators are widely used in
local analyses of tomographic estimators and uncertainty propagation
\cite{fessler1996meanvariance,engl1996regularization,stuart2010bayesian}.

Let
\begin{equation}
\delta\tilde{\mathbf{x}}^{(s)}
=
\tilde{\mathbf{x}}^{(s)}
-
\bar{\tilde{\mathbf{x}}}
\label{eq:sample_fluctuation_normalized}
\end{equation}
denote the normalized image fluctuation of sample \(s\). Applying
\(\mathbf{H}\) to this fluctuation gives its data-weighted
projection--backprojection response. We define the corresponding
data-visible variance as
\begin{equation}
\sigma^2_{\bm{H}\delta x}(j)
=
\left\langle
\left[
\bm{H}
\delta\tilde{\bm{x}}
\right]_j^2
\right\rangle ,
\qquad
\sigma_{\bm{H}\delta x}(j)
=
\sqrt{\sigma^2_{\bm{H}\delta x}(j)} .
\label{eq:dvv}
\end{equation}
The quantity \(\sigma_{\bm{H}\delta x}(j)\) will be referred to as the
\emph{data-visible standard deviation}.

One could also analyze the variance of the projected fluctuations
\(\bm{P}\delta\tilde{\bm{x}}\) directly in projection space. This would
provide a valid detector-space diagnostic, but it would not localize the
data-coupled variability in the reconstructed image domain. The weighted
projection--backprojection form used here maps the visible component of
the sampled fluctuations back into image space, where it can be compared
directly with reconstructed structures, residual backprojections, and
attenuation-model changes.

The motivation for this diagnostic is that \(\mathbf{H}\) suppresses
fluctuations associated with directions that are weakly constrained by
the measurements. In particular, for image-space modes satisfying
\(\mathbf{P}\mathbf{v}\approx 0\), one also has
\(\mathbf{H}\mathbf{v}\approx 0\). Therefore,
\(\sigma_{\bm{H}\delta x}\) emphasizes fluctuations that remain coupled to
measurable projection-space variations, rather than fluctuations confined
primarily to the null or near-null space of the imaging system.

This quantity should be interpreted as a data-coupled fluctuation
diagnostic, not as a replacement for the total image variance. The
ordinary sample variance \(\sigma^2(\tilde{x}_j)\) describes the full
variability of the sampled ensemble in image space, including components
that may be weakly visible to the data. In contrast,
\(\sigma_{\bm{H}\delta x}\) isolates the component of the sampled variability
that is propagated through the measurement operator under the assumed
forward model. Since \(\mathbf{H}\) inherits the numerical normalization
of \(\mathbf{P}\) and \(\mathbf{W}\), the absolute scale of
\(\sigma_{\bm{H}\delta x}\) should be interpreted for a fixed operator
normalization.

To quantify the influence of null and near-null directions, we
diagonalize \(\mathbf{H}\) in the masked image domain,
\begin{equation}
\mathbf{H}\mathbf{u}_k
=
\lambda_k\mathbf{u}_k .
\label{eq:H_eigenproblem}
\end{equation}
Modes with small \(\lambda_k\) correspond to image perturbations that
produce little change in the weighted projection data. A numerical null
or near-null subspace can therefore be defined by a cutoff
\(\lambda_{\rm cut}\). For each sampled fluctuation, we write
\begin{equation}
\delta\tilde{\mathbf{x}}^{(s)}
=
\delta\tilde{\mathbf{x}}^{(s)}_{>}
+
\delta\tilde{\mathbf{x}}^{(s)}_{<},
\label{eq:null_visible_decomp}
\end{equation}
where
\begin{equation}
\delta\tilde{\mathbf{x}}^{(s)}_{>}
=
\sum_{\lambda_k>\lambda_{\rm cut}}
a_k^{(s)}\mathbf{u}_k,
\qquad
\delta\tilde{\mathbf{x}}^{(s)}_{<}
=
\sum_{\lambda_k\leq\lambda_{\rm cut}}
a_k^{(s)}\mathbf{u}_k,
\qquad
a_k^{(s)}
=
\mathbf{u}_k^{\top}
\delta\tilde{\mathbf{x}}^{(s)} .
\label{eq:null_visible_components}
\end{equation}
This separates the ordinary image variance into contributions from modes
above and below the cutoff.

The same decomposition clarifies the behavior of the data-visible
variance. Since
\begin{equation}
\mathbf{H}
\delta\tilde{\mathbf{x}}^{(s)}
=
\sum_k
\lambda_k a_k^{(s)}\mathbf{u}_k,
\label{eq:H_eigenmode_weighting}
\end{equation}
the contribution of each mode to the data-visible fluctuation is weighted
by its eigenvalue. Consequently, low-\(\lambda_k\) modes may contribute
substantially to ordinary image variance, while contributing little to
the data-visible variance.

It is also useful to compare this ensemble-based diagnostic with the
backprojection of the weighted residual,
\begin{equation}
\mathbf{b}
=
\mathbf{P}^{\top}
\mathbf{W}
\left(
\mathbf{y}
-
\beta\mathbf{P}\bar{\tilde{\mathbf{x}}}
\right),
\label{eq:weighted_residual_backprojection}
\end{equation}
where \(\beta\) is a scalar normalization factor.
The quantity \(\mathbf{b}\) is a first-order diagnostic of model inconsistency or bias
in the ensemble mean. It is particularly sensitive to systematic mismatch
in the forward model, such as the omission or imperfect modelling of
attenuation correction. By contrast, the data-visible variance is a
second-order ensemble diagnostic: it measures the magnitude of sampled
image fluctuations after propagation through the same measurement
operator.

Both quantities are blind to the exact null space of the projection
operator. If \(\mathbf{P}\mathbf{v}=0\), then
\[
\mathbf{v}^{\top}\mathbf{b}
=
(\mathbf{P}\mathbf{v})^{\top}\mathbf{W}
\left(
\mathbf{y}
-
\beta\mathbf{P}\bar{\tilde{\mathbf{x}}}
\right)
=
0,
\]
and similarly
\[
\mathbf{v}^{\top}\mathbf{H}\mathbf{v}
=
(\mathbf{P}\mathbf{v})^{\top}
\mathbf{W}
(\mathbf{P}\mathbf{v})
=
0 .
\]
Thus, neither diagnostic is intended to quantify total null-space
uncertainty. Rather, both probe the data-visible subspace: the residual
backprojection identifies systematic mismatch of the ensemble mean,
whereas the data-visible variance provides the corresponding fluctuation
scale of the sampled ensemble in directions visible to the data.
\begingroup
\color{revisionblue}
The empirical spectral decomposition and attenuation-model comparison are
reported in Sec.~\ref{sec:empirical_dvv_validation}.
\endgroup

\subsection{Stochastic Gradient Descent}

Hamiltonian Monte Carlo sampling is computationally more demanding than
deterministic optimization. In practice, ensemble generation was therefore
initialized from an image obtained by minimizing the weighted
\(\chi^2\) objective using Adam, an adaptive stochastic-gradient descent
optimizer.

Within the present framework, this SGD-based reconstruction serves
primarily as an efficient initialization mechanism for the subsequent HMC
sampling. It provides a point estimate located near a high-probability
region of the sampling distribution, thereby reducing burn-in time and
improving sampling stability. The same optimization procedure can also be
used as a conventional point-estimate reconstruction; under matched
forward-model conditions it gives results globally comparable to those
obtained with MLEM.

The loss function used in the implementation was the mean weighted
squared projection residual,
\begin{equation}
\mathcal{L}_{\chi^2}
=
\frac{1}{N_{\rm bin}}
\sum_i
\frac{\big[y_i-(\bm{P}\bm{x})_i\big]^2}{y_i+\epsilon},
\label{eq:chi2loss}
\end{equation}
where \(N_{\rm bin}\) is the number of sinogram bins. This differs from
the \(\chi^2\) discrepancy introduced earlier only by a constant
normalization factor and therefore has the same minimizer. The constant
\(\epsilon\) prevents numerical instability in empty or near-empty bins.
After each optimizer update, the image was projected onto the
non-negative domain to ensure physically meaningful activity values.

In the present implementation, the forward model is represented by a
precomputed projection matrix \(\bm{P}\) for each reconstruction geometry.
During both SGD initialization and HMC sampling, \(\bm{P}\), the image
vector \(\bm{x}\) and the measured sinogram \(\bm{y}\) 
are stored as TensorFlow tensors. These operations are performed using
GPU-accelerated TensorFlow linear algebra when a compatible backend is
available, with no CPU--GPU transfer inside the reconstruction iterations
except for diagnostic output or final conversion for analysis and
plotting.

\subsection{Validation strategy}

To evaluate the proposed stochastic reconstruction framework, a hierarchical
validation strategy is adopted to distinguish algorithmic performance from
forward-model limitations and experimental uncertainties.

First, an idealized reconstruction scenario is considered in which projection
data are generated with effectively infinite counting statistics and an
idealized collimator model. In this regime the forward model closely matches
the data-generating process, allowing the stochastic implementation to be
benchmarked against deterministic iterative reconstruction methods. The
objective of this stage is to verify that the stochastic optimization and
sampling procedures converge to the same solution as established
deterministic approaches when the inverse problem is well defined.

Second, progressively more realistic software phantom simulations are
introduced, incorporating photon attenuation, geometric blur, and
collimator response. These studies allow systematic investigation of
forward-model adequacy and of the proposed data-visible ensemble
diagnostic.

Third, experimental measurements obtained using an anthropomorphic
neck--thyroid phantom are analyzed. These experiments provide realistic
acquisition statistics together with known activity distributions,
enabling quantitative validation of ensemble-based observables.

Finally, a clinical DATSCAN SPECT study is considered to illustrate the
applicability of the framework under practical imaging conditions where
ground truth is not available. In this case the ensemble is used primarily
to assess forward-model consistency and to estimate the variability of
clinically relevant derived quantities.

In addition to conventional global similarity measures, we employ two
visual diagnostic metrics introduced in~\cite{metrics}: spatially
resolved \(\chi^2\) maps and voxel-intensity histograms. Here the term
\(\chi^2\) map does not refer to the scalar projection-space objective
of Eq.~\eqref{eq:amias_chi2_diagonal}, but to a spatially resolved
diagnostic visualization of local mismatch. When a reference or ground
truth image is available, this map can be constructed directly in image
space. For experimental or clinical data, where ground truth is not
available, the diagnostic is instead based on the mismatch between the
measured and forward-projected data and its spatial backprojection.
Intensity histograms summarize the statistical distribution of
reconstructed voxel values and provide a complementary global assessment
of reconstruction consistency with the expected activity distribution.
Together, these metrics offer intuitive visual diagnostics that
complement the ensemble-based analysis presented in the following
sections.

\section{Results and Discussion}

\begingroup
\color{revisionblue}
The performance of the proposed stochastic reconstruction framework
is evaluated across a sequence of increasingly realistic scenarios.
We first report the empirical checks that support the statistical
approximation and the ensemble-based data-visible diagnostic introduced
in Sec.~2. We then consider the idealized software phantom benchmark,
realistic simulations incorporating attenuation and system effects,
experimental measurements obtained with an anthropomorphic thyroid
phantom, and finally a clinical DATSCAN SPECT study.
\endgroup

\begingroup
\color{revisionblue}
\subsection{Empirical validation of the statistical approximation and sampled ensemble}
\label{sec:empirical_probability_validation}

The central-limit test in Fig.~\ref{fig:amias_probability_validation}(a)
was performed to justify the Gaussian compatibility approximation for
the detector-bin observables used in the reconstruction. For this test we
used a fully three-dimensional voxelized Shepp--Logan phantom defined on
a \(128\times128\times128\) Cartesian grid. The corresponding SPECT
acquisition was simulated with detailed photon-transport physics,
including attenuation, scattering, gamma-camera response, and collimator
modelling. A set of 500 statistically independent realizations was
generated under identical imaging conditions and used to examine the
distribution of detector-bin counts and their block averages.

The independent realizations were grouped into blocks, and the average
counts in a representative detector bin were examined. As expected from
the central limit theorem, the block-average distributions become
narrower and approach the Gaussian forms associated with the counting
variance. This supports the use of the weighted least-squares form in
Eq.~\eqref{eq:amias_chi2_diagonal} for the count regime considered here.
The test does not establish that all detector bins are statistically
independent; it only checks the Gaussian counting approximation
underlying the adopted data-compatibility measure. Further details of the
software phantom and simulation setup are given in
Sec.~\ref{subsec:software_phantoms}.

An empirical consistency check of the sampled image ensemble is shown in
Fig.~\ref{fig:amias_probability_validation}(b--d). This test used the same three-dimensional voxelized Shepp--Logan phantom
and GATE simulation setup as the central-limit test in panel~(a),
including the same detailed photon-transport physics and detector
modelling. For the image-domain validation, however, 100 independent noisy acquisitions were generated,
each with a total detected count level of approximately \(10^6\). Each
realization was reconstructed using the same reconstruction protocol,
thereby providing an empirical distribution of reconstructed images under
repeated statistically independent measurements.

This repeated-simulation ensemble was then compared with the HMC ensemble
generated from one representative acquisition from the same simulation
setup. The comparison was performed using circular regions rather than
individual voxels. The region radius was chosen from the effective
point-spread-function radius estimated for the same reconstruction
pipeline. This is necessary because individual voxel values are strongly
affected by the system and reconstruction response and should not be
interpreted as independent resolution elements. PSF-sized regional
averages provide a more meaningful scale for comparing reconstructed
activity fluctuations.

The sampled ensemble reproduces the dominant spread of the repeated
reconstruction experiment for the selected regional observables. 
This is illustrated in Fig.~\ref{fig:amias_probability_validation}(b),
which shows the ratio of the HMC ensemble standard deviation 
to that obtained from repeated simulations for representative background, low-, medium-, and high-activity regions. Representative regional distributions
are shown in Fig.~\ref{fig:amias_probability_validation}(c). 
A complementary projection-space check is shown in
Fig.~\ref{fig:amias_probability_validation}(d), where the 
repeated simulated sinograms are compared with the
predictive ones obtained from the sampled image ensemble after
forward prediction with \(\bm{P}\bm{x}\).

These comparisons are not intended as a formal proof that the sampled
ensemble is an exact posterior distribution. Rather, they provide an
empirical consistency check showing that the adopted compatibility density
and sampling procedure reproduce the dominant regional image-space and projection-space
variability for data generated under the
same acquisition model.

\begin{figure}[th!]
\centering
\includegraphics[width=\linewidth]{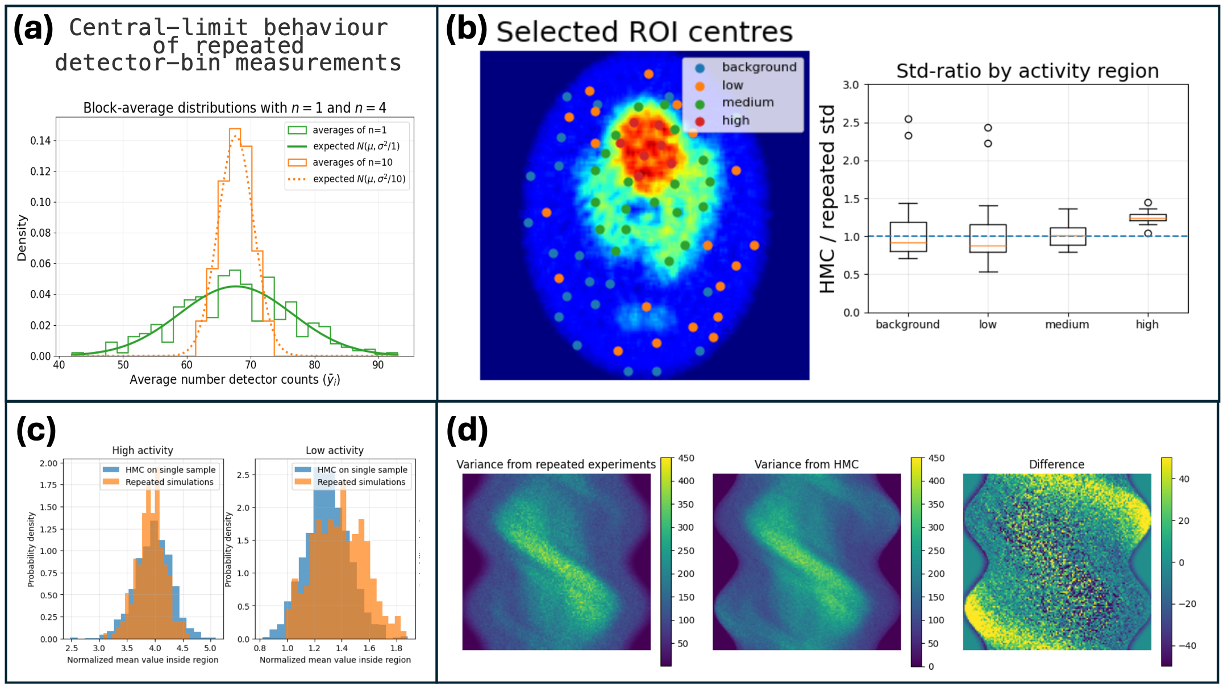}
\caption{
Empirical validation of the image ensemble generated from the
likelihood-based compatibility density. (a) Central-limit behaviour of
repeated detector-bin measurements. Independent simulations of the same
object were grouped into blocks, and the block-average count
distributions were compared with the corresponding Gaussian
approximations. This test supports the Gaussian weighted least-squares
compatibility factor used in Eq.~\eqref{eq:amias_chi2_diagonal} for the
count regime considered here. (b) Regional image-domain comparison.
Circular regions were selected across background, low-, medium-, and
high-activity parts of the reconstructed image. The radius of each region
was chosen from the effective point-spread-function radius of the
reconstruction pipeline, so that the comparison is performed at the
spatial scale of the reconstructed resolution rather than at the level of
individual voxels. For each region, the standard deviation of the
normalized regional density \(\rho_R\) estimated from the sampled image
ensemble was divided by the corresponding value obtained from repeated
independent simulations followed by reconstruction. Values close to unity
indicate good agreement in the predicted regional variability.
(c) Representative distributions of \(\rho_R\) for high- and low-activity
regions, comparing the sampled ensemble from a representative acquisition
with the ensemble obtained from repeated simulations.
(d) Projection-space validation. The standard deviation estimated from
repeated simulated sinogram realizations is compared with the
corresponding projection-space prediction obtained by forward projection
of the sampled image ensemble. The difference image shows the residual
between the two estimates. Further details of the simulations are given
in Sec.~\ref{subsec:software_phantoms}, and HMC implementation details
are provided in Appendix~\ref{app:hmc_implementation}.
}
\label{fig:amias_probability_validation}
\end{figure}
\endgroup

\begingroup
\color{revisionblue}
\subsection{Empirical assessment of the data-visible variance diagnostic}
\label{sec:empirical_dvv_validation}

\begin{figure}[t]
\centering
\includegraphics[width=\linewidth]{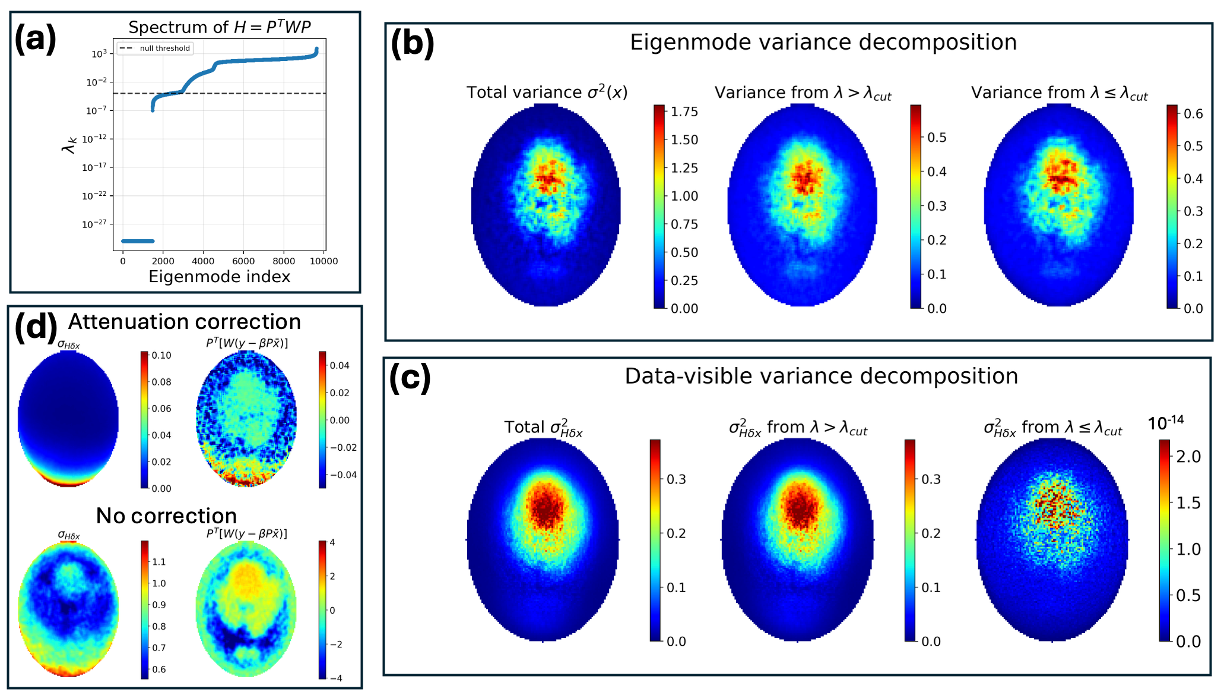}
\caption{
Data-visible variance, null/near-null-space decomposition, and comparison
with the weighted residual backprojection. (a) Spectral analysis of
\(\mathbf{H}=\mathbf{P}^{\top}\mathbf{W}\mathbf{P}\) in the masked image
domain. 
% The histogram of \(\log_{10}\lambda_k\) shows a low-eigenvalue
% sector separated from the main spectral bulk. 
The dashed line indicates
the cutoff \(\lambda_{\rm cut}\) used to distinguish weakly constrained
modes from the remaining subspace. (b) Decomposition of the ordinary
image variance. The total variance \(\sigma^2(\tilde{x})\) is separated
into contributions from modes with \(\lambda>\lambda_{\rm cut}\) and
\(\lambda\leq\lambda_{\rm cut}\). The low-eigenvalue modes contribute
substantially to the ordinary image variance, showing that the sampled
ensemble contains fluctuations in weakly constrained directions. (c)
Decomposition of the data-visible standard deviation. The contribution
from low-eigenvalue modes is strongly suppressed because
\(\mathbf{H}\mathbf{u}_k=\lambda_k\mathbf{u}_k\). Thus, modes that
contribute appreciably to ordinary image variance but are weakly coupled
to the measured projection data are largely removed from the
data-visible variance. (d) Comparison between the data-visible standard
deviation and the weighted residual backprojection
\(\mathbf{P}^{\top}\mathbf{W}
(\mathbf{y}-\beta\mathbf{P}\bar{\tilde{\mathbf{x}}})\)
for reconstructions with and without attenuation correction. The residual
backprojection remains sensitive to structured mismatch in the ensemble
mean, whereas the data-visible variance measures the scale of sampled
fluctuations that remain coupled to the data. With attenuation correction,
the data-visible variance is strongly reduced, while the residual
structure is also reduced but not completely eliminated.
}
\label{fig:data_visible_nullspace_summary}
\end{figure}

Figure~\ref{fig:data_visible_nullspace_summary} summarizes this analysis
for a representative reconstructed image ensemble. The ordinary sample
variance contains both data-visible and weakly constrained components.
In contrast, the data-visible variance emphasizes the component of the
sampled variability that remains coupled to the measurement process. The
spectral decomposition shows that low-eigenvalue modes can contribute
appreciably to the ordinary image variance, whereas their contribution to
the data-visible variance is strongly suppressed by the action of
\(\mathbf{H}\).

Figure~\ref{fig:data_visible_nullspace_summary}(d) illustrates this
distinction using reconstructions with and without attenuation
correction. Omitting attenuation correction produces a strong structured
weighted-residual backprojection, as expected for a forward-model
mismatch. The data-visible variance is also increased, indicating larger
data-coupled fluctuations in the sampled ensemble. With attenuation
correction, the data-visible variance becomes very small, while the
backprojected residual remains structured but is substantially reduced.
This behaviour indicates that the two diagnostics measure different
aspects of the reconstruction. A small data-visible variance suggests
that, within the assumed forward model and sampled distribution, little
statistical ensemble variability remains in directions visible to the
projection data. If a structured residual persists at the same time, the
remaining mismatch is more naturally associated with systematic effects,
model bias, or limitations of the forward model, rather than with
additional statistically sampled image fluctuations.

This interpretation is empirical and should not be read as a proof that
no further improvement is possible. Instead, it provides a practical
criterion for separating two regimes: data-visible statistical
variability, which is captured by the sampled ensemble, and residual
structured mismatch, which points to possible systematic limitations of
the reconstruction model. This comparison motivates the use of the
data-visible standard deviation in the following sections as an
ensemble-based measure of whether changes in the forward model produce
additional data-coupled changes in the reconstructed image ensemble.
\endgroup

\subsection{Reconstruction parameters and convergence control}
\label{sec:reconstruction_parameters}

All reconstructions used fixed convergence criteria rather than iteration
numbers selected by visual inspection. The numerical parameters were used
only to reach a stable solution or stationary sampling regime, not to
optimize image appearance.

For HMC, samples contributing to image means, variances, and data-visible
standard-deviation maps were collected only after thermalization. The
leapfrog parameters, acceptance monitoring, parallel tempering,
autocorrelation analysis, thinning, and criteria for obtaining effectively
uncorrelated samples are described in
Appendix~\ref{app:hmc_implementation}. These parameters affect sampling
efficiency and thermalization, whereas the reported statistics use only
post-equilibration samples.

MLEM was initialized with a uniform image inside the active field of view
and updated using the standard multiplicative rule with sensitivity
normalization. Positivity was enforced, pixels outside the active field
were masked, and a small constant \(\epsilon\) stabilized division in
empty or near-empty projection bins. Convergence was monitored through
the weighted mean-square discrepancy between measured and predicted
projections. Automatic stopping occurred when its relative change
remained below \(10^{-5}\) for five consecutive iterations after at least
five iterations.

The SGD reconstructions used for initialization and comparison minimized
the weighted \(\chi^2\) objective in Eq.~\eqref{eq:chi2loss} using Adam,
an adaptive stochastic-gradient descent method, with initial learning
rate \(0.1\). The learning rate was reduced by a factor of \(0.99\) when
the loss did not improve over the prescribed patience interval, and the
image was projected onto the non-negative domain after each update.
After at least 50 iterations, optimization stopped when the relative loss
improvement remained below \(10^{-5}\) for 30 consecutive iterations.
The image with the lowest loss was retained.

Thus, MLEM and SGD results were taken after convergence, while HMC
observables were computed from post-equilibration samples, ensuring that
the comparisons reflect differences in the forward model and ensemble
statistics rather than arbitrary iteration counts.

\subsection{Software Phantoms}
\label{subsec:software_phantoms}

% ===== BEGIN SUBSTANTIVE REVISION =====
% \begingroup
% \color{revisionblue}
Before moving into experimental data, we used a controlled simulation
environment to assess the applicability and effectiveness of the
stochastic reconstruction implementation. The simulation study was
performed using the GATE10 Monte Carlo toolkit~\cite{GATE2022}, which
enables detailed modelling of particle transport and detector response in
nuclear medicine imaging. The simulated scanner was based on a dual-head
SPECT gamma-camera geometry with an explicit collimator model, so that
the camera and collimator response were included directly in the Monte
Carlo simulation.

We used a three-dimensional Shepp--Logan-like voxelized phantom with
realistic background activity. The phantom was constructed from
ellipsoidal regions, each assigned a specified activity, and is available
in the project repository.\footnote{\url{https://github.com/leontiou/PERSPECT/tree/main/GATE10/SheppLogan3D}}
The phantom was defined on a \(128\times128\times128\) Cartesian voxel
grid, with an isotropic voxel size of \(3.125~\mathrm{mm}\). The primary
emission energy was set to 159~keV, corresponding to the main photon
energy of \(^{123}\mathrm{I}\).
\begin{itemize}
    \item Primary sources: 12~MBq
    \item Intermediate sources: 325~kBq to 1~MBq
    \item Low-uptake regions: 10--40~kBq
\end{itemize}

The acquisition consisted of $120$ projection views, and the
resulting sinogram contained a total of \(3.28\times10^{6}\) detected
counts. The simulation included photon attenuation, scattering, detector
response, and collimator effects, providing a realistic test case for
evaluating reconstruction performance under controlled conditions. The
output sinogram generated by this simulation was used as input for both
Hamiltonian Monte Carlo and gradient-based reconstruction methods.

\subsubsection{Idealized benchmark against deterministic reconstruction}

As a first validation step, we benchmark the stochastic reconstruction
framework in a controlled limiting regime by defining an
\emph{Ideal Image} and corresponding \emph{Ideal Sinogram} for the
Shepp--Logan software phantom. Following the standardized
image taxonomy proposed in~\cite{metrics}, the Ideal Image is not
identical to the Source phantom; rather, it represents the best
physically achievable reconstruction under asymptotic acquisition
conditions.

In this configuration, the forward model retains the irreducible physics
of photon transport, while avoidable instrumental and statistical
limitations are strongly suppressed. This is achieved by using an
effectively unphysical very-high-count acquisition together with an ideal
collimator assumption, so that finite counting noise, detector blur, and
collimator-induced degradation are minimized. This case should therefore
be interpreted as a limiting benchmark rather than as a clinically
realistic acquisition. Under these conditions, remaining discrepancies
mainly reflect the forward model and intrinsic transport effects rather
than statistical noise.

A deterministic gradient-based \(\chi^2\) reconstruction was first used
to locate a high-probability region of the image space. This image was
then used as the initial configuration for Hamiltonian Monte Carlo
sampling. The HMC reconstruction reported here is the ensemble mean of the
post-thermalization samples. For comparison, a Maximum Likelihood
Expectation Maximization (MLEM) reconstruction was also performed under
the same idealized forward-model assumptions.

\begin{figure*}[th]
\centering
\includegraphics[width=0.9\textwidth]{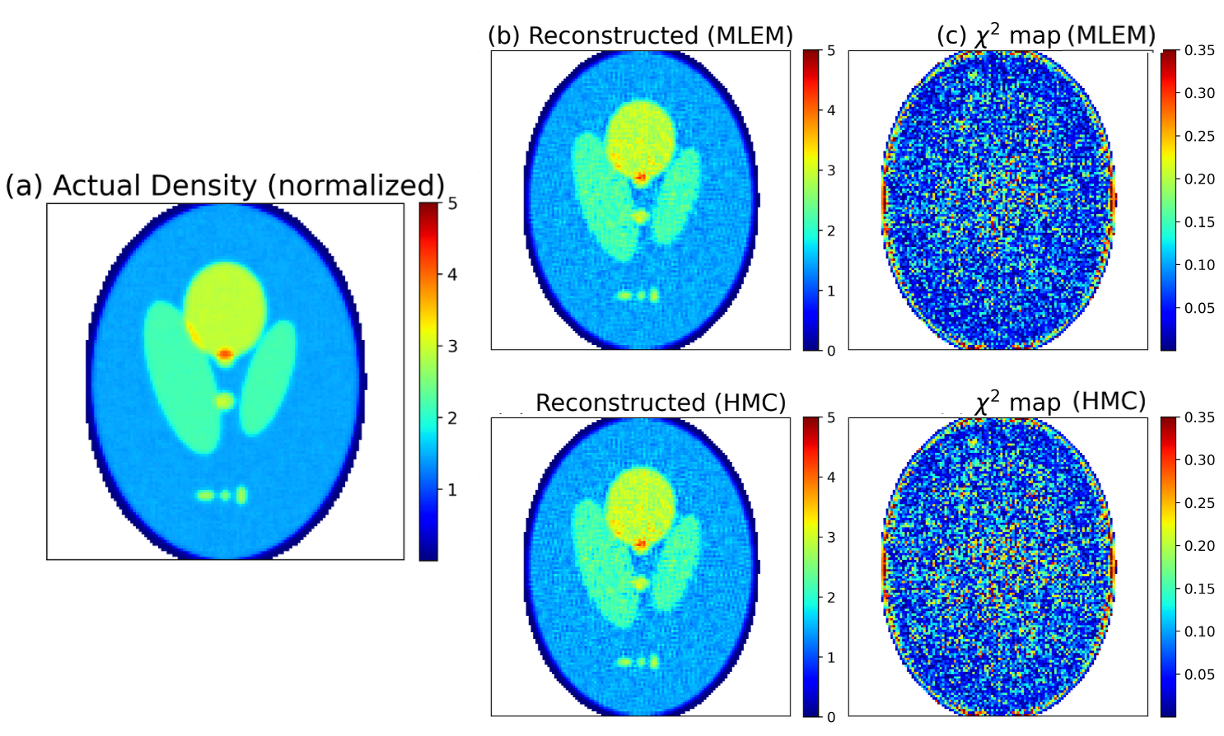}
\caption{
Idealized benchmark against deterministic reconstruction.
(a) True activity distribution, normalized to the same scale as the
reconstructions. The right panels show the MLEM reconstruction and local
\(\chi^2\) map, together with the HMC ensemble-mean reconstruction and
its corresponding local \(\chi^2\) map. This benchmark uses an
effectively unphysical very-high-count acquisition and an ideal collimator
assumption, and should therefore be interpreted as a limiting consistency
test rather than as a realistic imaging scenario. The \(\chi^2\) map
metric is analyzed in detail in~~\protect\cite{metrics}.
}
\label{fig:sgd}
\end{figure*}

The resulting HMC ensemble mean and corresponding local \(\chi^2\) map
are shown in Fig.~\ref{fig:sgd}. The comparison with MLEM is used here as
a deterministic reference for the limiting reconstruction problem. The
purpose is not to claim identical local residual structure, but to verify
that the HMC ensemble mean remains globally consistent with a standard
deterministic reconstruction when the forward model is matched and the
counting statistics are effectively asymptotic.

\begin{figure}[th]
\centering
\includegraphics[width=0.5\textwidth]{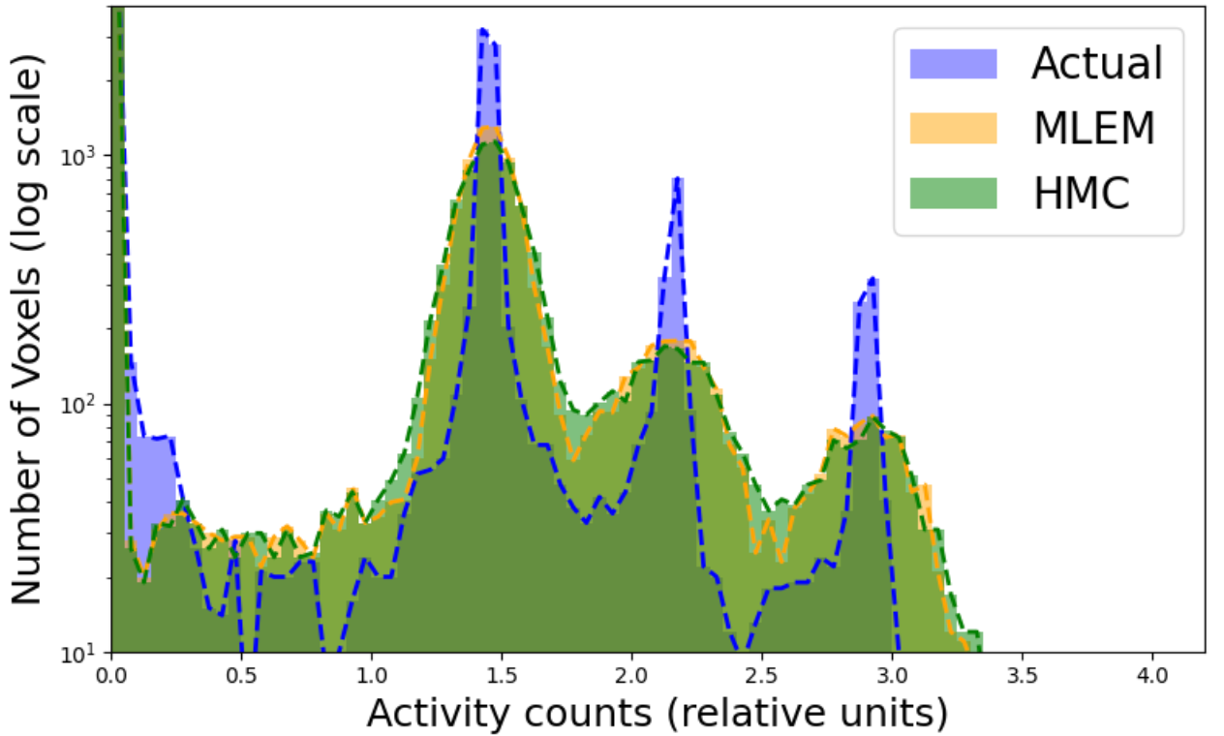}
\caption{
Voxel-intensity histograms for the true image, MLEM reconstruction, and
HMC ensemble mean. The comparison shows global agreement in the recovered
intensity distribution under idealized high-count conditions. This metric
is analyzed in detail in~~\protect\cite{metrics}.
}
\label{fig:histograms}
\end{figure}

The voxel-intensity histograms in Fig.~\ref{fig:histograms} compare the
true image, MLEM reconstruction, and HMC ensemble mean. The histograms and
global image-quality metrics in Table~\ref{tab:metrics} indicate that the HMC mean and MLEM produce
globally comparable reconstructions in this limiting case. 

\begin{table}[h]
\centering
\caption{Reconstruction quality metrics for the idealized benchmark
(mean \(\pm\) standard deviation).}
\begin{tabular}{|l|c|c|}
\hline
\textbf{Metric} & \textbf{MLEM} & \textbf{HMC ensemble mean} \\
\hline
MSE  & \(0.01039 \pm 0.00065\) & \(0.01299 \pm 0.00075\) \\
MAE  & \(0.06266 \pm 0.00155\) & \(0.06964 \pm 0.00167\) \\
PSNR & \(29.82 \pm 0.01\)     & \(29.74745 \pm 0.01394\) \\
SSIM & \(0.99358 \pm 0.00781\)& \(0.99198 \pm 0.00780\) \\
\hline
\end{tabular}
\label{tab:metrics}
\end{table}

These results establish that, in the asymptotic high-count limit with a
matched idealized forward model, the HMC ensemble mean is globally
consistent with standard deterministic reconstructions. The stochastic
formulation is therefore not introduced to improve the point estimate in
this idealized regime. Its purpose is to provide access to an ensemble of
data-compatible images around the reconstructed solution, enabling the
uncertainty and data-visible fluctuation analyses that become relevant in
the realistic acquisition studies below.
% \endgroup
% ===== END SUBSTANTIVE REVISION =====

\subsubsection{Realistic simulation with attenuation and scatter}

The mean reconstructions obtained with Hamiltonian Monte Carlo (HMC)
are comparable to those produced by stochastic gradient descent (SGD)
and deterministic MLEM methods. The primary objective of HMC in the
present work, however, is not improved point estimation but the
extraction of physically interpretable fluctuation structure from the
generated ensemble.

HMC produces an ensemble of data-compatible reconstructed images sampled
from the adopted likelihood-based compatibility density proportional to
\(e^{-\chi^2/2}\). Each sample
represents a plausible activity distribution compatible with the
measured data and the assumed forward model. Expectation values of
arbitrary observables can therefore be evaluated directly as ensemble
averages.

To evaluate the framework under realistic imaging conditions,
we constructed a fully three-dimensional voxelized software phantom
and simulated the complete SPECT acquisition chain using the GATE
Monte Carlo toolkit. Forward projections were generated with full
photon-transport modeling, including attenuation and collimator
effects. This three-dimensional validation is essential because
subsequent sections use voxelized attenuation maps derived from
experimental measurements, requiring methodological consistency.

Unlike the idealized benchmark, the present scenario incorporates
attenuation, geometric blur, and realistic collimator response.
Reconstruction accuracy is therefore reduced relative to the ideal
case. The goal here is not to optimize image quality per se, but to
examine whether ensemble-based diagnostics can distinguish intrinsic
physics limitations from forward-model deficiencies.

Three attenuation configurations were considered:

\begin{itemize}
    \item \textbf{Model 1:} uniform attenuation with an ideal collimator assumption.
    \item \textbf{Model 2:} attenuation correction combined with a realistic collimator model.
    \item \textbf{Model 3:} incomplete attenuation correction using an exponentially decreasing attenuation coefficient.
\end{itemize}

Introducing uniform attenuation correction (Model~1) substantially
reduces the large-scale structured mismatch observed without attenuation
correction. This reduction is visible both in the spatially resolved
\(\chi^2\) map and in the weighted residual backprojection
\(\mathbf{P}^{T}\mathbf{W}(\mathbf{y}-\mathbf{P}\bar{\mathbf{x}})\).
With further refinement in Model~2, which incorporates realistic
collimator response, the residual structure is further suppressed and
the remaining \(\chi^2\) pattern becomes largely noise-like.

The data-visible standard deviation
\(\sigma_{\bm{H}\delta x}\) provides complementary information. From NC to
Model~1 and from Model~1 to Model~2, it decreases markedly, indicating
that the sampled ensemble contains progressively less variability in
directions that are visible to the projection data. In this sense,
\(\sigma_{\bm{H}\delta x}\) is the more appropriate diagnostic for assessing
the statistical content of the reconstructed ensemble. By contrast, the
backprojected residual remains a diagnostic of the ensemble mean and is
therefore more sensitive to residual bias or systematic mismatch.

Model~3 intentionally introduces attenuation mismatch through a
depth-dependent under-correction. Although the mean reconstruction remains
visually similar to those of Models~1--2, the \(\chi^2\) map and the
backprojected residual show renewed structured mismatch. The
data-visible standard deviation also increases relative to Model~2 and
regains a spatial gradient. This confirms that the ensemble fluctuations
again project into data-sensitive directions when the forward model is
made inconsistent.

The combined interpretation is therefore that the residual backprojection
and the data-visible variance should not be treated as competing metrics.
The residual identifies structured mismatch of the mean image, whereas
\(\sigma_{\bm{H}\delta x}\) quantifies the remaining ensemble variability
visible to the measured data. In the attenuation-corrected cases, the
data-visible variance becomes small even when some residual structure
persists, suggesting that further improvement is more likely associated
with systematic modelling effects than with additional statistically
sampled image fluctuations under the same forward model.

\begin{figure*}[th!]
\centering
\includegraphics[width=\textwidth]{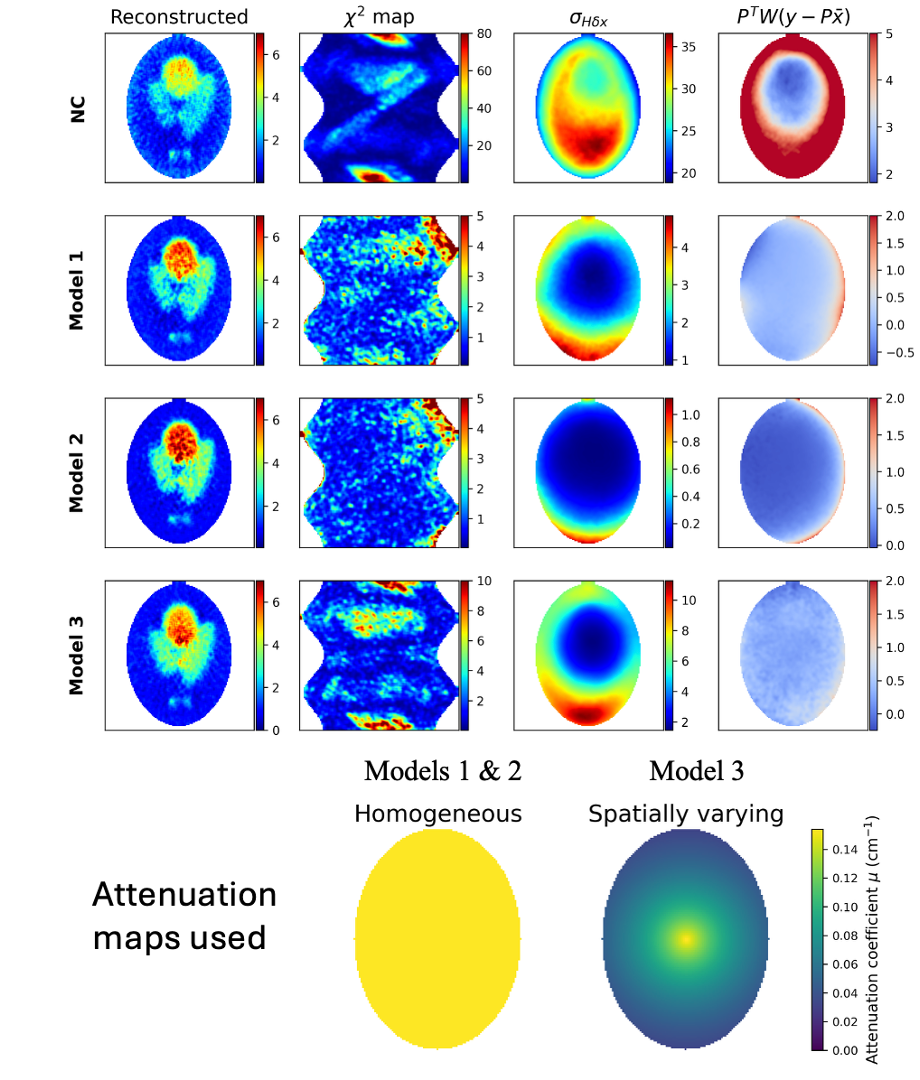}
\caption{
Application of the HMC framework to a realistic 3D voxelized
software phantom. The upper rows show, for each forward-model
configuration, the ensemble-mean reconstruction, the spatially resolved
\(\chi^2\) map, the sampled data-visible standard deviation
\(\sigma_{\bm{H}\delta x}\), and the weighted residual backprojection
\(\bm{P}^{T}\bm{W}(\bm{y}-\bm{P}\bar{\bm{x}})\). Rows correspond to
NC, Model 1, Model 2, and Model 3. The bottom row shows the attenuation
coefficient maps used in the attenuation models: a homogeneous
water-equivalent map and a spatially varying map, both displayed with a
common colorbar in units of \(\mathrm{cm}^{-1}\).
}
\label{fig:software_hmc}
\end{figure*}

These effects cannot be inferred from reconstructed images alone.
The ensemble-based HMC framework therefore provides a practical
diagnostic tool for separating physics-imposed limitations from
forward-model inadequacy through spatial analysis of ensemble
fluctuations.

Up to this point, the evaluation has focused primarily on qualitative
diagnostics derived from voxel-wise residual and variance maps.
In the following subsection we extend the analysis to quantitative
measurements extracted directly from the ensemble. By evaluating
physically meaningful observables—such as lesion-to-lesion activity
ratios—together with their associated uncertainties, we demonstrate
how ensemble-based reconstruction enables statistically grounded
model comparison beyond visual assessment alone.

\subsection{Application to an Anthropomorphic Phantom}

To demonstrate the practical relevance of the proposed Hamiltonian
Monte Carlo (HMC)--based reconstruction and uncertainty quantification
framework, we apply it to an anthropomorphic neck--thyroid phantom
designed for postsurgical SPECT/CT imaging. The phantom, introduced
and characterized in detail by 
Michael \textit{et al.}~\cite{phantom1,phantom2},
% \citeasnoun{phantom1,phantom2},
incorporates anatomically realistic structures and tissue-equivalent materials.

The phantom consists of a modular neck geometry manufactured using a
combination of 3D printing and molding techniques. It includes thyroid
remnant inserts with known volumes ranging from approximately $0.5$~mL
to $10$~mL, positioned at clinically realistic locations. Surrounding
anatomical structures such as the trachea, oesophagus, cervical spine,
and clavicles are also represented. The activity distribution within
the phantom can be controlled to emulate realistic
background-to-remnant uptake ratios, allowing systematic investigation
of small-volume quantification under challenging imaging conditions.

This design enables controlled validation of reconstruction accuracy
in the presence of partial-volume effects, limited spatial resolution,
photon attenuation, and scatter, which are known to degrade
conventional deterministic volume estimation methods. In the present
work we analyze one representative \(^{123}\mathrm{I}\) background
phantom acquisition from this dataset, acquired on a dual-head GE Infinia
Hawkeye 4 SPECT/CT system with LEHR collimators, 60 projection views, a
\(128\times128\) projection matrix, and a processed sinogram total of
\(4.14\times10^{5}\) detected counts.

\subsubsection{Forward Model Hierarchy and Diagnostic Objective}

The experimental acquisitions were used primarily to demonstrate how
HMC can serve as a diagnostic tool for forward-model assessment,
rather than to optimize reconstruction accuracy. The objective is to
evaluate how successive refinements of the forward model affect the
consistency between measured projections and reconstructed images,
and to determine the spatial origin of the remaining uncertainty.

We consider three forward modeling configurations of increasing
physical realism:

\begin{enumerate}
    \item \textbf{Geometric model (NC):} a purely geometric projection
    operator that neglects photon attenuation.

    \item \textbf{GATE homogeneous attenuation model (Model 1):}
    attenuation correction implemented in GATE assuming a homogeneous
    water-equivalent attenuation medium. This model includes detailed
    photon transport and collimator response but does not account for
    spatial variations in tissue attenuation.

    \item \textbf{Macroscopic non-uniform attenuation model (Model 2):}
    a computationally efficient forward model implemented in Python,
    incorporating spatially varying attenuation coefficients derived
    from the experimentally provided attenuation map. This model
    includes macroscopic photoelectric absorption and Compton
    scattering but does not explicitly simulate collimator physics.
    This complementary modeling structure allows partial disentanglement
    of transport realism (collimator response) from spatial attenuation
    heterogeneity, although neither model is strictly complete.
\end{enumerate}

The GATE-based model therefore improves physical realism through
accurate transport and collimator modeling but assumes spatially
uniform attenuation. In contrast, the macroscopic model introduces
spatial heterogeneity in attenuation while remaining computationally
lightweight. This complementary modeling structure allows us to
separate the impact of transport realism from that of spatial
attenuation heterogeneity.

\begin{figure}[th!]
    \centering
    \includegraphics[width=0.45\textwidth]{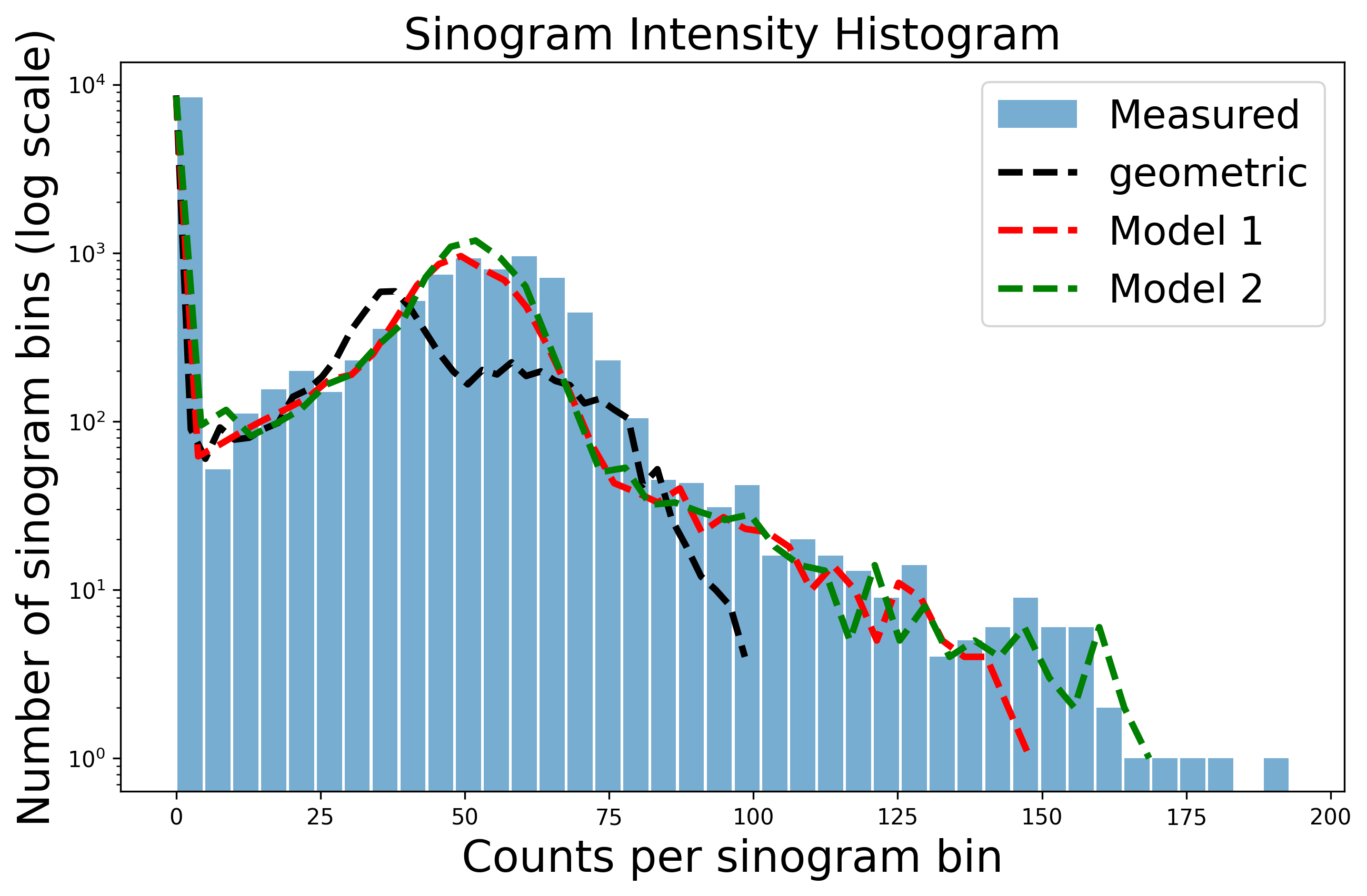}
\caption{
Sinogram intensity histograms comparing measured projection data with
projections predicted by different forward models. The purely geometric
model (black dashed line) shows substantial discrepancy with the
measured data. Introducing attenuation correction using a homogeneous
water-equivalent medium in GATE (red dashed line, Model~1) improves
agreement across a broad range of count levels. Further refinement is
observed when a non-uniform attenuation model derived from the
experimental attenuation map is used (green dashed line, Model~2).
While these histograms demonstrate progressive global improvement,
they do not reveal the spatial origin of the remaining discrepancies.
}
\label{fig:sinogram_hist}
\end{figure}

Figure~\ref{fig:sinogram_hist} shows that global agreement between
measured and predicted sinograms improves as model complexity
increases. However, histogram-based metrics remain intrinsically
global and cannot determine whether improvements affect
data-constrained components of the solution or merely alter weakly
constrained directions of the inverse problem.

\subsubsection{HMC-Based Spatial Diagnostic}

\begin{figure*}[th!]
    \centering
    \includegraphics[width=0.95\textwidth]{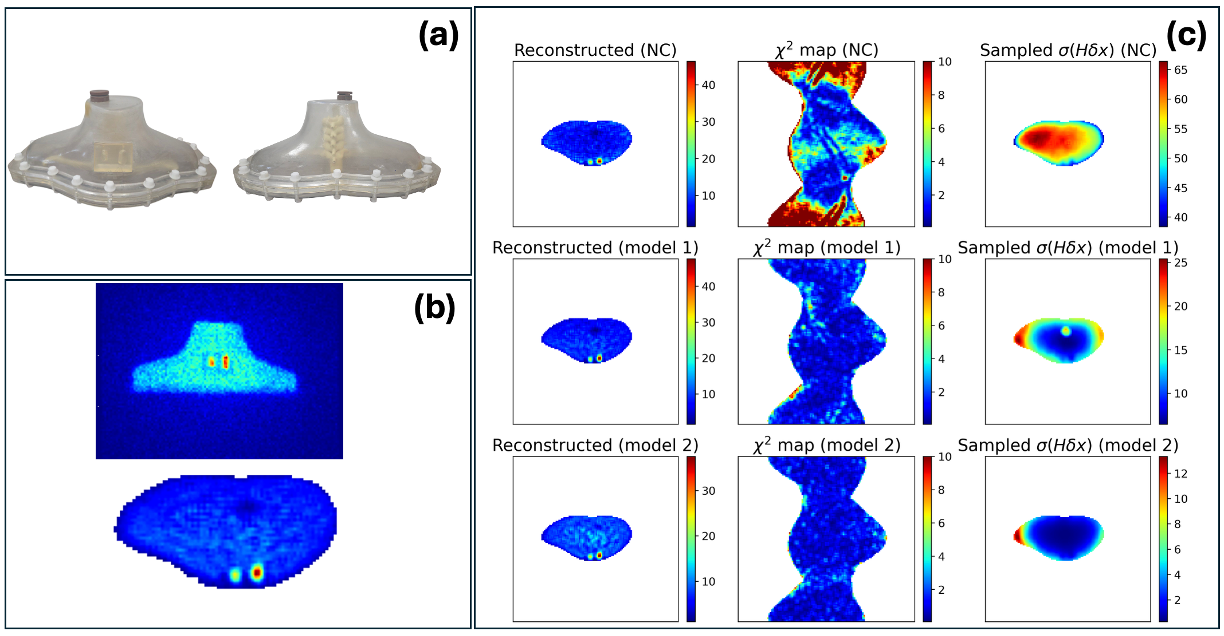}
    \caption{
Application of the HMC framework to the thyroid phantom experiment.
Panel (a) shows photographs of the experimental thyroid phantom.
Panel (b) displays a representative detector projection (top) and the
corresponding reconstructed central slice (bottom) used for the analysis.
Panel (c) presents the reconstruction analysis.
Each row corresponds to a different forward model.
Left column: reconstructed central slice (ensemble mean).
Middle column: spatially resolved $\chi^2$ map comparing measured and predicted sinograms.
Right column: sampled data-visible deviation $\sigma_{\bm{H}\delta x}$.\\
Top row (NC): geometric projection model without attenuation correction.
Second row (Model 1): GATE-based simulation using a homogeneous
water-equivalent attenuation model and collimator response.
Third row (Model 2): macroscopic non-uniform attenuation model
with spatially varying attenuation coefficients optimized from the experimental data.\\
While global $\chi^2$ decreases with progressive model refinement,
the $\chi^2$ maps become increasingly noise-like, limiting their spatial
diagnostic value. In contrast, the sampled $\sigma_{\bm{H}\delta x}$ maps
provide a clearer indication of how image fluctuations project into
data-sensitive directions. The non-uniform attenuation model produces
a marked suppression of $\sigma_{\bm{H}\delta x}$ near object boundaries,
demonstrating that forward-model improvements primarily affect
regions where attenuation heterogeneity is most significant,
while the bulk variability remains largely governed by intrinsic
ill-posedness and counting statistics. The $\sigma_{\bm{H}\delta x}$ maps are
shown normalized by the mean reconstructed activity within the object
mask (dimensionless, relative to image mean).
}
\label{fig:thyroid_hmc}
\end{figure*}

To obtain spatially resolved insight, HMC sampling is applied to each
forward-model configuration. The resulting ensemble enables computation of voxel-wise variability and of the data-visible standard deviation
\(\sigma_{\bm{H}\delta x}\), which quantifies how strongly image fluctuations
project into data-sensitive directions defined by the measurement operator.

As shown in Fig.~\ref{fig:thyroid_hmc}, the geometric model produces
structured $\chi^2$ residuals, indicating significant forward-model
mismatch. The introduction of homogeneous attenuation correction in
GATE reduces these discrepancies, although residual structure
remains, consistent with unmodeled spatial heterogeneity.

The macroscopic non-uniform attenuation model further suppresses
structured residuals, particularly near boundaries and in regions where
attenuation gradients are strongest. A notable feature is the reduction of
the data-visible variance in the central region corresponding
approximately to the cervical spine. This region is poorly represented by
the homogeneous attenuation assumption used in Model~1, because the
central bony structure introduces attenuation heterogeneity that is not
captured by a water-equivalent medium. The higher
\(\sigma_{\bm{H}\delta x}\) observed for Model~1 in this region therefore
indicates that the sampled image fluctuations remain coupled to the data
through an inadequately modelled attenuation structure. In Model~2, the
use of a non-uniform attenuation map reduces this central data-visible
component, suggesting improved consistency between the measured
projections and the attenuation model.

For the refined model, \(\sigma_{\bm{H}\delta x}\) is markedly reduced in
regions where attenuation modeling is most critical, while the remaining
variability in the bulk appears comparatively more uniform. This behavior
indicates that residual variability is increasingly dominated by
intrinsically weakly constrained modes of the inverse problem rather than
by the specific attenuation-model inadequacy visible in Model~1.

Such distinctions cannot be extracted from reconstructed images,
$\chi^2$ maps, or global sinogram statistics alone.
The ensemble-based HMC framework therefore provides a physically
interpretable separation between modeling error and fundamental
inverse-problem limitations, offering a practical tool for
forward-model validation in experimental imaging studies.

\subsubsection{Quantitative lesion-to-lesion ratio (thyroid phantom)}
\label{sec:thyroid_ratio}

To complement the visual residual and variance maps, we extracted a simple quantitative
observable from the image ensemble: the ratio of mean reconstructed intensity
between the two hot inserts of the thyroid phantom. For the acquisition ID analyzed here
(ID~8 in Table~1 of \cite{phantom1}), the two lesions were prepared with approximately
equal activity concentration ($A_1/V_1 \approx A_2/V_2$), hence the expected concentration
ratio is $R_{\mathrm{true}}\approx 1$. The lesion sizes are close to the intrinsic spatial
resolution of SPECT, placing the measurement near the partial-volume and resolution limits
of the system.

Once the ensemble $\{x^{(s)}\}_{s=1}^{N_s}$ is available, any observable that can be
expressed as a function of the reconstructed image may be evaluated sample-wise and
assigned an uncertainty directly from the ensemble distribution. The reliability of this
uncertainty estimate depends on the adequacy of the forward model and the statistical
convergence of the sampling process.

For each attenuation-model configuration (NC, Model~1, Model~2), two circular ROIs were
defined around the lesion locations on each axial slice. The ROI radius $r$ was varied
to assess robustness with respect to ROI definition. For a fixed radius $r$, the mean
ROI intensity for lesion $\ell\in\{1,2\}$ was estimated per sample as a variance-weighted
average,

\begin{equation}
\mu_\ell^{(s)}(r)=
\frac{\sum_{j\in \Omega_\ell(r)} w_j\, x_j^{(s)}}
     {\sum_{j\in \Omega_\ell(r)} w_j},
\qquad
w_j=\frac{1}{\widehat{\mathrm{Var}}(x_j)+\epsilon},
\label{eq:roi_weighted_mean}
\end{equation}

where $\Omega_\ell(r)$ denotes the set of voxels inside the circular ROI
(aggregated over the included axial slices), $\widehat{\mathrm{Var}}(x_j)$ is the
sample variance estimated from the ensemble, and $\epsilon$ prevents numerical
instability.

The lesion-to-lesion ratio was then computed for each sample,

\begin{equation}
R^{(s)}(r)=\frac{\mu_2^{(s)}(r)}{\mu_1^{(s)}(r)},
\end{equation}

and summarized by the mean and standard deviation,

\begin{equation}
\bar R(r)=\langle R^{(s)}(r)\rangle,
\qquad
\sigma_R(r)=\mathrm{std}\!\left(R^{(s)}(r)\right).
\label{eq:R_mean_std}
\end{equation}

This sample-wise ratio automatically incorporates covariance between the
two ROI means, since it is evaluated prior to ensemble averaging.

The specific ROI construction adopted here is not essential; alternative
segmentation strategies (e.g.\ threshold-based or Otsu-type methods)
could be used equivalently. The purpose of this example is not to introduce
a new segmentation methodology, but to demonstrate that once an ensemble
is available, any chosen observable can be computed together with a
well-defined uncertainty.

\begin{figure}[ht]
\centering
\includegraphics[width=0.95\linewidth]{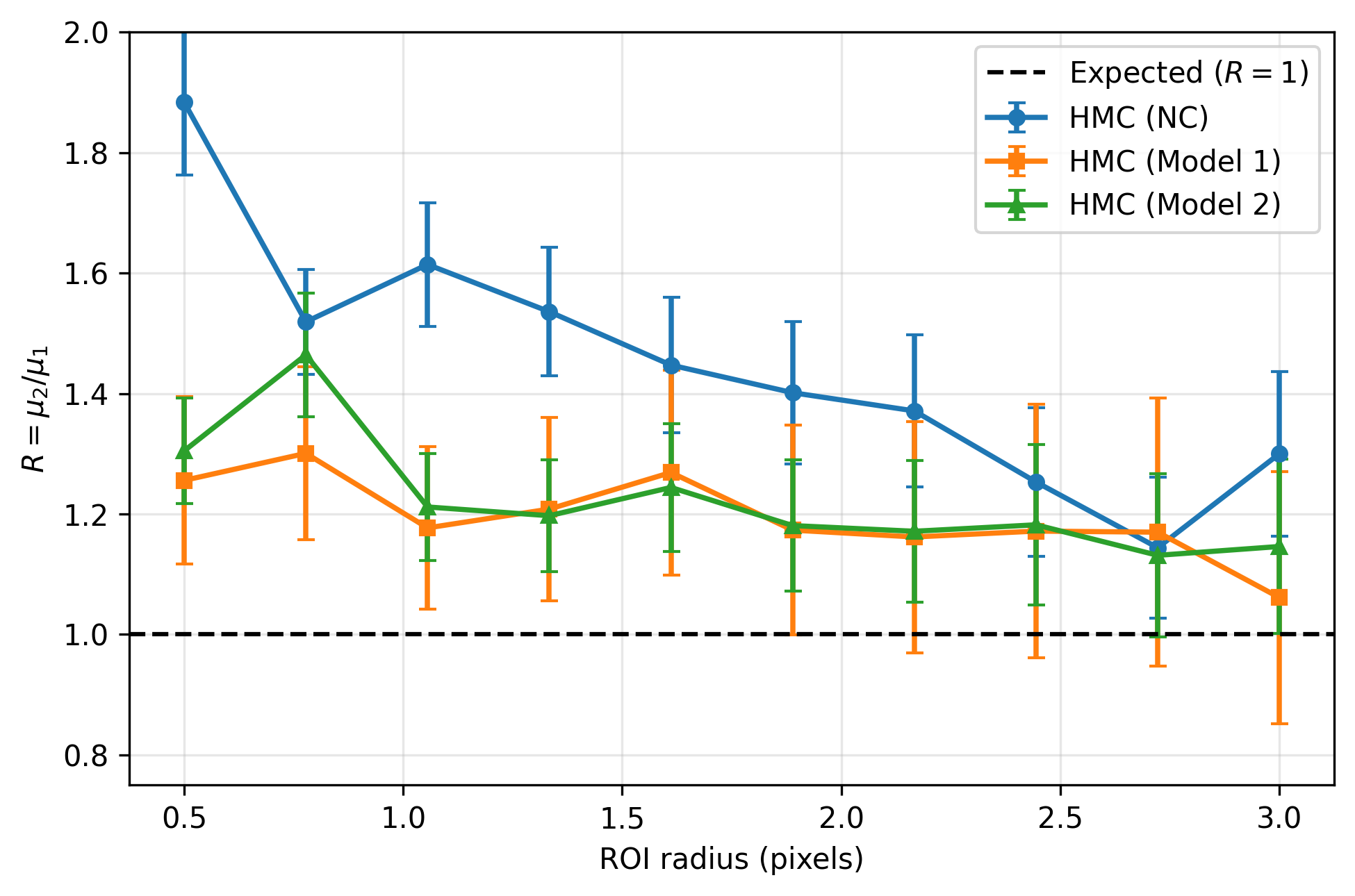}
\caption{
Ensemble-based quantitative assessment from the thyroid phantom.
For each ROI radius $r$, the lesion-to-lesion mean-intensity ratio
$R(r)=\mu_2(r)/\mu_1(r)$ was computed sample-wise from the HMC ensemble
using variance-weighted ROI means (Eq.~\ref{eq:roi_weighted_mean}) and
summarized by the mean and standard deviation (Eq.~\ref{eq:R_mean_std}).
Error bars represent the standard deviation of $R(r)$.
The dashed line indicates the expected concentration ratio
$R_{\mathrm{true}}=1$.
The plateau behaviour demonstrates robustness with respect to ROI radius,
while convergence toward $R_{\mathrm{true}}$ and reduced inter-model spread
with improved attenuation modeling indicate enhanced quantitative
consistency. This example illustrates how ensemble-based reconstruction
enables estimation of derived observables together with ensemble-derived 
uncertainty estimates under the adopted sampling model.
}
\label{fig:thyroid_ratio}
\end{figure}

Figure~\ref{fig:thyroid_ratio} shows \(\bar R(r)\) as a function of ROI
radius for the three forward-model configurations. The estimates exhibit
a plateau over a broad range of radii, indicating stability with respect
to ROI choice and demonstrating that the measurement does not rely on
ad hoc thresholding. The recovered ratio remains biased with respect to
the expected value \(R_{\mathrm{true}}\approx1\). This bias is attributed
primarily to partial-volume effects and limited SPECT spatial resolution,
since the lesion dimensions are close to the intrinsic resolution of the
system. Additional contributions may arise from imperfect attenuation
modelling, ROI placement, and residual mismatch between the reconstruction
model and the experimental acquisition.

The purpose of this analysis is not to introduce a partial-volume-corrected
estimator, but to demonstrate how a derived observable can be evaluated
sample-wise from the HMC ensemble. In principle, the bias could be reduced
by incorporating recovery coefficients, resolution modelling, or
calibration-based partial-volume correction. Such corrections were not
applied here in order to keep the observable directly tied to the sampled
reconstruction ensemble.

The spatial variance used in Eq.~\ref{eq:roi_weighted_mean} and the ratio
estimates in Eqs.~\ref{eq:R_mean_std} are computed from the same retained
post-thermalization HMC samples. The variance map is first estimated from
the ensemble and used to define fixed voxel weights within each ROI. The
lesion means and their ratio are then evaluated sample-wise using these
fixed weights, so that the reported uncertainty reflects the distribution
of the derived observable across the same image ensemble.

% \endgroup
% ===== END SUBSTANTIVE REVISION =====

\subsection{Application to clinical data}

To illustrate the applicability of the proposed framework to clinical
data, we processed a DATSCAN SPECT study (I-123 radiopharmaceutical)
of a patient diagnosed with Parkinson’s disease, for which no
CT-based attenuation map was available. In the absence of
patient-specific attenuation information, a preliminary reconstruction
was first performed without attenuation correction (NC).
Subsequently, homogeneous attenuation models were introduced to
approximate photon attenuation effects, including photoelectric
absorption and Compton scattering.

% ===== BEGIN SUBSTANTIVE REVISION =====
% \begingroup
% \color{revisionblue}
The clinical DATSCAN acquisition was performed on a Segami Mirage SPECT
system. The processed projection data consisted of 128 projection views
with a \(128\times128\) matrix and pixel spacing of
\(3.95\times3.95~\mathrm{mm}^2\). The total number of counts in the
processed sinogram was \(2.65\times10^{6}\).
% \endgroup
% ===== END SUBSTANTIVE REVISION =====

Previous investigations have examined the role of attenuation correction (AC) in 
I-123-ioflupane (DATSCAN) SPECT, particularly in relation to quantitative accuracy 
and diagnostic performance. CT-based AC has been shown to systematically modify 
striatal uptake ratios and reduce bias associated with photon attenuation in the 
skull and soft tissues \cite{Lange2014}. While such corrections improve quantitative 
consistency and anatomical plausibility, their impact on visual diagnostic 
classification is generally moderate. Similarly, Bieńkiewicz et al.\ reported that 
CT-based AC alters quantitative binding ratios and may slightly improve diagnostic 
discrimination between extrapyramidal syndromes, although the effect on overall 
clinical categorization remains limited \cite{Bienkiewicz2008}. More recent 
quantitative workflows employing automated threshold-based analysis tools, such 
as DaTQUANT, further indicate that attenuation correction contributes to improved 
reproducibility and calibration of quantitative metrics when standardized uptake 
thresholds are applied \cite{Neill2021}. 

Collectively, these studies suggest that attenuation correction enhances quantitative 
fidelity but does not necessarily produce large changes in clinical classification. 
In this context, the clinical DATSCAN study presented here is not intended to evaluate 
diagnostic performance per se, but rather to demonstrate how an ensemble-based 
Hamiltonian Monte Carlo (HMC) reconstruction framework can provide a 
physics-grounded assessment of forward-model adequacy. By analyzing image 
fluctuations and their projection into data-sensitive directions through the 
operator \(\bm{H}=\bm{P}^{\top}\bm{W}\bm{P}\), the proposed approach allows evaluation of whether 
successive attenuation refinements meaningfully reduce data-constrained variability, 
thereby offering a complementary perspective beyond conventional residual or 
histogram-based comparisons.

\begin{figure*}[th!]
    \centering
    \includegraphics[width=0.95\textwidth]{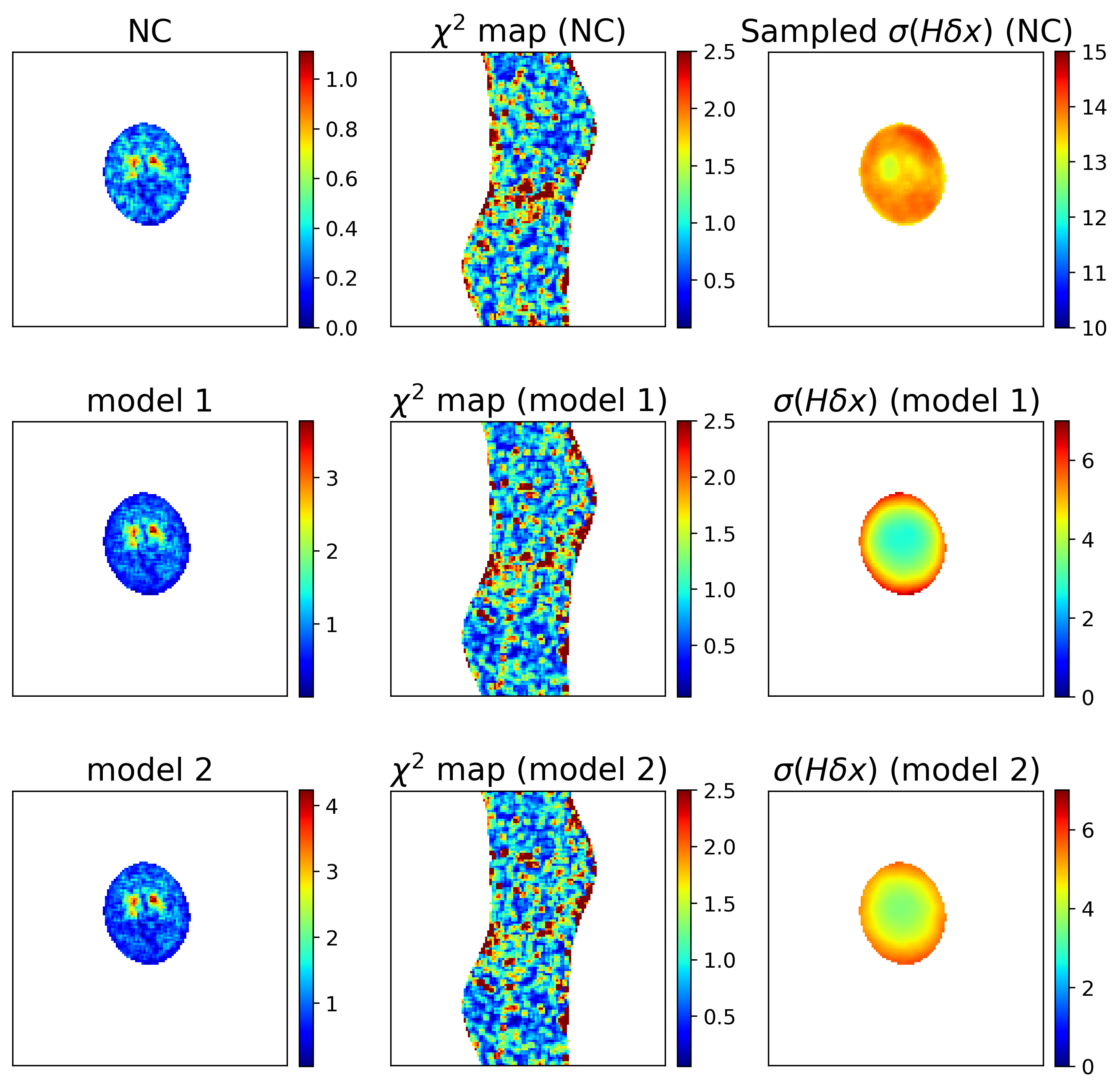}
    \caption{
Application of the HMC framework to a clinical DATSCAN SPECT study.
Each row corresponds to a different forward-model configuration.
Left column: reconstructed central slice (ensemble mean).
Middle column: spatially resolved $\chi^2$ map comparing measured and
predicted sinograms.
Right column: sampled data-visible standard deviation $\sigma_{\bm{H}\delta x}$.
Top row (NC): geometric projection without attenuation correction.
Middle row (Model~1): homogeneous water-equivalent attenuation model.
Bottom row (Model~2): homogeneous water-equivalent attenuation model
augmented with a skull component.
While the $\chi^2$ maps exhibit largely noise-like residual structure
with limited distinction between models, the $\sigma_{\bm{H}\delta x}$ maps
provide a clearer spatial diagnostic of variability.
The absence of systematic suppression of $\sigma_{\bm{H}\delta x}$ across
these attenuation refinements suggests that further improvement
requires investigation of additional forward-model components beyond
simple homogeneous attenuation modeling.
The higher overall magnitude of $\sigma_{\bm{H}\delta x}$ compared with the
controlled phantom studies reflects the absence of patient-specific
attenuation information and the increased modeling uncertainty
inherent to clinical data.
}
\label{fig:datscan_hmc}
\end{figure*}

In addition to the purely geometric projection model (NC), two
attenuation configurations were examined.
Model~1 assumes a homogeneous water-equivalent attenuation medium.
Model~2 augments this homogeneous model by including a skull component,
thereby introducing a simple two-region attenuation description.

At the sinogram level, as qualitatively indicated by the spatially
resolved $\chi^2$ maps (Fig.~\ref{fig:datscan_hmc}), the differences
between the attenuation configurations are subtle and largely
noise-like. In such clinical settings, where ground truth is not
available and residual maps provide limited spatial guidance, it is
difficult to determine whether modifications of the attenuation model
lead to meaningful improvements in the reconstruction.

The sampled data-visible standard deviation $\sigma_{\bm{H}\delta x}$ provides a more
structured diagnostic. The introduction of a homogeneous attenuation
model modifies the spatial distribution of $\sigma_{\bm{H}\delta x}$
relative to the purely geometric case. However, further refinement
within the class of simple homogeneous attenuation models does not
produce a systematic reduction of this quantity. This behavior
suggests that the dominant sources of variability in this clinical
dataset are not controlled primarily by attenuation scaling alone,
but instead arise from other components of the imaging chain.

In this sense, the ensemble-based analysis remains informative even
when this standard deviation does not decrease substantially: it indicates that
further refinement of homogeneous attenuation models yields limited
benefit and that improvements may require investigation of additional
forward-model components, such as patient-specific attenuation maps,
depth-dependent collimator response, scatter modeling beyond simple
approximations, detector resolution effects, or improved noise models.
The proposed ensemble-based metrics therefore provide a practical means
of identifying when refinement of a particular forward-model component
yields diminishing returns in clinical reconstructions.

\subsubsection{Quantitative striatal binding ratio from ensembles (clinical DATSCAN)}
\label{sec:datscan_sbr}

To complement the qualitative residual and variance diagnostics of
Fig.~\ref{fig:datscan_hmc}, we extracted a clinically relevant
quantitative observable from the image ensemble: the striatal binding
ratio (SBR), defined relative to a posterior occipital reference region.
DATSCAN acquisitions are frequently assessed using striatal-to-reference
uptake ratios to summarize dopaminergic transporter binding, and the
present analysis uses this ratio as an illustrative example of
ensemble-based quantification under low-count clinical conditions.

Given an HMC ensemble $\{x^{(s)}\}_{s=1}^{N_s}$ and binary masks for the
left striatum $\Omega_L$, right striatum $\Omega_R$, and reference region
$\Omega_{\mathrm{ref}}$, we compute sample-wise mean intensities

\begin{equation}
\mu_k^{(s)} =
\frac{1}{|\Omega_k|}
\sum_{j\in\Omega_k} x_j^{(s)},
\qquad
k\in\{L,R,\mathrm{ref}\},
\end{equation}

and the corresponding sample-wise binding ratios

\begin{equation}
\mathrm{SBR}_L^{(s)}=
\frac{\mu_L^{(s)}-\mu_{\mathrm{ref}}^{(s)}}{\mu_{\mathrm{ref}}^{(s)}},
\qquad
\mathrm{SBR}_R^{(s)}=
\frac{\mu_R^{(s)}-\mu_{\mathrm{ref}}^{(s)}}{\mu_{\mathrm{ref}}^{(s)}}.
\label{eq:sbr_def}
\end{equation}

The point estimate and uncertainty are then obtained directly from the
ensemble:

\begin{equation}
\overline{\mathrm{SBR}}_k=\langle \mathrm{SBR}_k^{(s)}\rangle,\qquad
\sigma_{\mathrm{SBR},k}=
\mathrm{std}\!\left(\mathrm{SBR}_k^{(s)}\right),
\qquad k\in\{L,R\}.
\end{equation}

This procedure illustrates a general principle: once an ensemble of
reconstructed images is available, any image-derived observable can be
evaluated sample-wise and reported with an uncertainty estimated from
the ensemble distribution. The reliability of this uncertainty depends
on the adequacy of the assumed forward model and the convergence of the
sampling process.

\begin{figure*}[th!]
\centering
\includegraphics[width=0.95\textwidth]{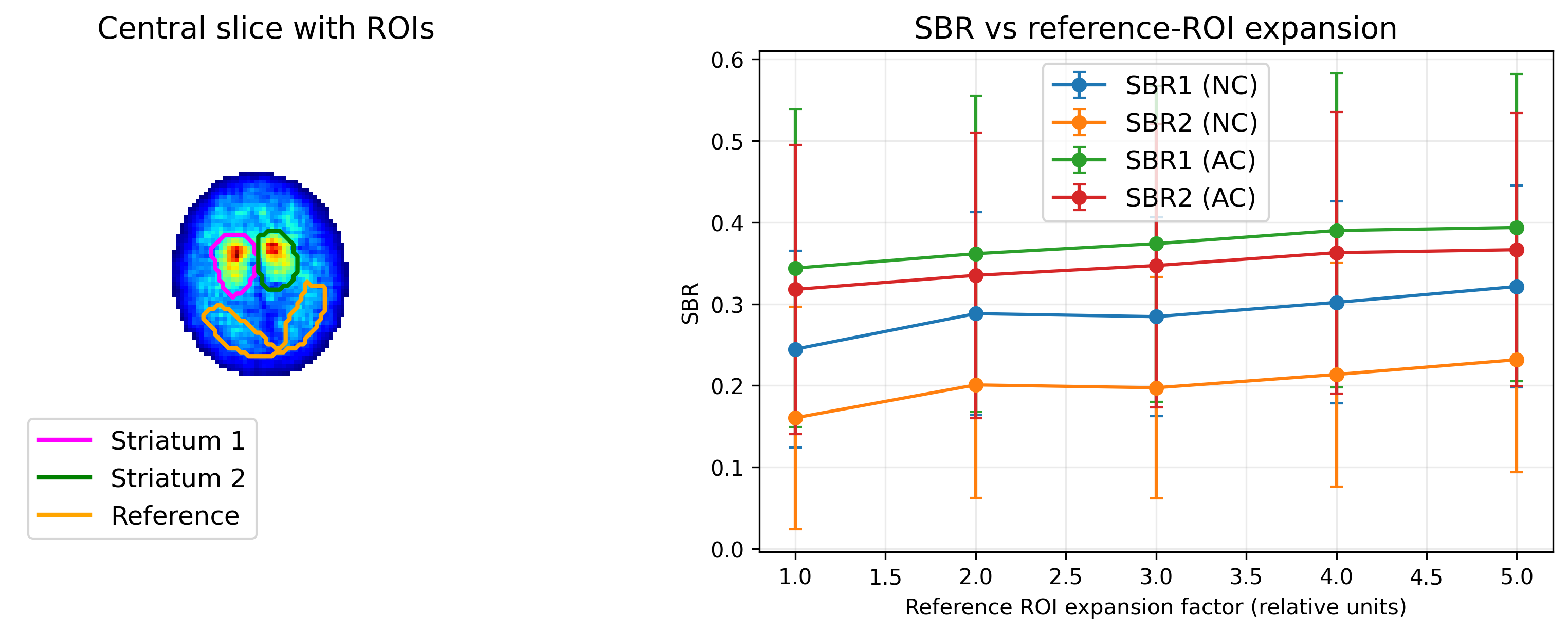}
\caption{
Quantitative striatal binding ratio (SBR) analysis for the clinical
DATSCAN study.
\textbf{Left:} central reconstructed slice (ensemble mean) with manually
defined regions of interest (ROIs). The two upper ROIs correspond to the
left and right striatal uptake regions, while the lower posterior ROI
defines the reference region used for SBR normalization.
\textbf{Right:} SBR estimates as a function of the fractional expansion
of the reference ROI used to test robustness with respect to ROI
definition. For each configuration, SBR values are computed sample-wise
from the HMC ensemble according to Eq.~(\ref{eq:sbr_def}), and the error
bars represent the standard deviation across the sampled images.
Results are shown for reconstructions without attenuation correction
(NC) and with homogeneous attenuation correction (AC). While the
absolute SBR values depend moderately on the reference ROI size, the
estimates remain relatively stable across a broad range of expansion
factors. The substantial uncertainty bars reflect the low-count
clinical acquisition and the absence of patient-specific attenuation
information. This example illustrates how ensemble-based reconstruction
enables reporting clinically relevant uptake ratios together with
uncertainty estimates derived directly from the sampled image ensemble.
}
\label{fig:datscan_sbr}
\end{figure*}

To assess robustness with respect to ROI definition, the reference ROI
was systematically expanded by a fractional factor while remaining
constrained to the brain mask, and the SBR was recomputed for each
expansion. Figure~\ref{fig:datscan_sbr} shows the resulting SBR trends
for two forward-model configurations (NC and AC). In this low-count
clinical setting, the uncertainty bars remain substantial, reflecting
the combined effects of counting statistics, limited spatial resolution,
and inverse-problem ambiguity. Nevertheless, the ensemble-based approach
provides quantitative estimates with principled error bars, enabling
interpretable model-to-model comparisons and supporting downstream
population analyses where uncertainty propagation is essential.

\subsection{Discussion}

The results presented in this work demonstrate that the stochastic reconstruction
framework based on Hamiltonian Monte Carlo (HMC) sampling can be applied
practically to voxel-based emission tomography problems and can provide
information that is not accessible through conventional point-estimate
reconstruction methods. While deterministic approaches such as MLEM or
gradient-based optimization aim to produce a single reconstructed image,
the ensemble generated by HMC represents a statistical description of the
set of images compatible with the measured projection data and the assumed
forward model.

% ===== BEGIN SUBSTANTIVE REVISION =====
% \begingroup
% \color{revisionblue}
Under controlled conditions, the HMC ensemble mean was found to be
globally comparable to the deterministic MLEM reconstruction. In the
idealized software-phantom benchmark, both methods produced similar
voxel-intensity distributions and comparable global image-quality
metrics. Local residual structures were not claimed to be identical.
In this limiting regime, the principal role of HMC is therefore not to
improve the point estimate, but to characterize the distribution of
data-compatible images around the reconstructed solution.
% \endgroup
% ===== END SUBSTANTIVE REVISION =====

The ensemble structure becomes particularly informative when realistic
acquisition conditions and modeling imperfections are introduced.
The sampled data-visible standard deviation, $\sigma_{\bm{H}\delta x}$, provides a spatially
resolved diagnostic that quantifies how fluctuations propagate
through the data-weighted projection--backprojection operator
\(\bm{H} = \bm{P}^{\top}\bm{W}\bm{P}\). This quantity isolates the component of uncertainty that
remains coupled to the measured data and therefore provides insight into
the conditioning of the inverse problem under the assumed forward model.
Across the software phantom and experimental phantom studies, improvements
in the forward model were reflected not only in reduced residual structure
in the $\chi^2$ maps but also in a systematic reduction and homogenization
of $\sigma_{\bm{H}\delta x}$ within the object. This behavior indicates that
fluctuations increasingly project into weakly constrained
directions of the inverse problem rather than into data-sensitive modes.

An important observation is that the spatial structure of
$\sigma_{\bm{H}\delta x}$ can reveal forward-model deficiencies even when
reconstructed images or residual maps appear visually acceptable.
In the controlled simulation where an intentionally incomplete attenuation
model was introduced, the mean reconstruction remained visually
similar to that obtained with more accurate models, while the
$\sigma_{\bm{H}\delta x}$ maps exhibited clear spatial structure reflecting the
model mismatch. This demonstrates that ensemble-based diagnostics can
provide complementary information beyond conventional reconstruction
evaluation metrics.

The application to the anthropomorphic thyroid phantom further illustrates
the practical use of image ensembles for quantitative measurements.
Because observables can be evaluated sample-wise over the ensemble,
derived quantities such as lesion-to-lesion activity ratios can be
estimated together with 
model-conditioned uncertainty estimates derived directly from the ensemble distribution.
Importantly, these uncertainty estimates
incorporate both counting statistics and the ambiguity of the inverse
problem, avoiding the need for ad hoc error propagation methods.

The clinical DATSCAN example highlights the challenges associated with
forward-model assessment in realistic imaging scenarios where ground truth
is not available. In this case the ensemble diagnostics indicate that
simple homogeneous attenuation corrections produce only limited reduction
of data-constrained uncertainty, suggesting that additional components of
the imaging model—such as patient-specific attenuation maps, depth-dependent
collimator response, or improved scatter modeling—may be required to
meaningfully improve reconstruction consistency.

A full calibration study of the ensemble-derived uncertainty estimates
for derived observables remains an important direction for future work.
Such a study would require repeated-noise experiments over multiple count
levels, object geometries, forward-model assumptions, and ROI definitions.
The uncertainty estimates reported here should therefore be interpreted as
model-conditioned ensemble uncertainties under the adopted sampling
density, supported by the empirical consistency checks of Fig.~\ref{fig:amias_probability_validation},
rather than as universally calibrated confidence intervals.

Despite these advantages, the proposed stochastic framework remains
computationally more demanding than conventional deterministic methods.
Hamiltonian Monte Carlo requires repeated evaluation of forward and
adjoint projections during the simulation of Hamiltonian trajectories,
which increases computational cost relative to standard iterative
reconstruction. However, the use of GPU-accelerated numerical libraries
and gradient-based initialization significantly reduces the practical
cost of ensemble generation, making the approach feasible on modern
workstations.

Overall, the results indicate that ensemble-based stochastic reconstruction
can complement existing deterministic reconstruction methods by providing
a physically interpretable framework for uncertainty quantification and
forward-model validation. Rather than replacing conventional algorithms,
the approach can serve as an additional diagnostic layer capable of
identifying the origin of residual uncertainty and guiding improvements
in forward-model formulation for emission tomography.

\section{Conclusions and Outlook}

In this work we presented a stochastic reconstruction framework for
emission tomography in which Hamiltonian Monte Carlo (HMC) sampling is
used to generate ensembles of data-compatible reconstructed images.
The HMC sampling stage is initialized using stochastic-gradient descent,
which provides a fast approximation of the reconstruction solution and
maintains computational practicality on standard GPU-enabled workstations.

Under controlled conditions, the stochastic reconstruction achieves
image quality comparable to established deterministic iterative
methods. The principal advantage of the approach, however, lies not
in marginal improvements of point estimates but in the information
contained in the ensemble of reconstructed images.
The ensemble allows direct evaluation of uncertainty and provides
spatially resolved diagnostics of the adequacy of the forward model.

In particular, we demonstrated that the sampled data-visible standard deviation
\(\sigma_{\bm{H}\delta x}\) provides a physically interpretable measure
of model adequacy. Through software phantom studies, anthropomorphic
phantom experiments, and a clinical DATSCAN acquisition, the ensemble
analysis helps distinguish uncertainty associated with the intrinsic
ill-posedness of the inverse problem from variability linked to
forward-model limitations. In practical terms, the method can
indicate whether refinement of a specific component of the imaging
model, such as attenuation or collimator modeling, produces a
meaningful reduction of data-constrained uncertainty.

The ensemble representation also enables estimation of clinically
relevant derived observables together with model-conditioned uncertainty
estimates obtained from the sampled image distribution.
In the DATSCAN
example presented here, the ensemble was used to estimate striatal
binding ratios (SBR) and their associated uncertainties under
low-count clinical conditions. Such uncertainty estimates can
complement existing quantitative metrics and may improve the
robustness of statistical analyses in population-based imaging studies.

More broadly, this work illustrates how modern GPU-enabled numerical
frameworks make ensemble-based stochastic reconstruction practically
accessible alongside traditional iterative methods such as MLEM.
By combining fast stochastic-gradient optimization with Hamiltonian
sampling, the proposed approach provides a practical path toward
uncertainty-aware emission tomography and physics-informed model
validation in realistic imaging scenarios.

\section*{Acknowledgements}
%\ack{
This work was supported by the Cyprus Research and Innovation 
Foundation under the project PERSPECT (Project No. EXCELLENCE/0524/0410). 
PERSPECT project is implemented in the frames of Cohesion Policy Programme 
“THALIA 2021-2027” and is co-funded by the European Union. 
In addition, this work was supported by computing time awarded 
on the Cyclone supercomputer of the High Performance Computing 
Facility of The Cyprus Institute under project ``Medical Imaging'' (mi).
%}

\section*{Data availability statement}

The data supporting the findings of this study consist of (i) simulated
phantom data generated by the authors and (ii) previously published and
anonymized clinical imaging data. A subset of the simulated data, including
the three-dimensional Shepp--Logan phantom and the anthropomorphic phantom
sinograms and attenuation maps, is publicly available through the PERSPECT
project GitHub repository:
\url{https://github.com/leontiou/PERSPECT}. The repository also provides
the methodology used to generate the projection matrices with GATE,
incorporating attenuation and scattering physics. Additional simulated
data are available from the corresponding author upon reasonable request.

\appendix

\section{Practical implementation of the HMC reconstruction}
\label{app:hmc_implementation}

% ===== BEGIN SUBSTANTIVE REVISION =====
% \begingroup
% \color{revisionblue}
The HMC ensembles were generated from the image-space compatibility
density defined in Eq.~\eqref{eq:amias_sampling_measure}. Chains were
initialized from the deterministic gradient-based reconstruction in order
to reduce thermalization time. The leapfrog step size was adapted during
warm-up to maintain a stable Metropolis acceptance rate, typically close
to \(70\%\). The number of leapfrog steps was usually \(L=10\)--\(20\),
and different stable choices of mass scale and trajectory length gave
equivalent ensemble statistics after proper thermalization.

Non-negativity was enforced during proposal generation. The reported
reconstructions used reflective boundary conditions at \(x_j=0\), while
projection and clipping of negative values were also tested as robustness
checks. These alternatives did not visibly alter the reconstructed image
quality or the main ensemble diagnostics.

Parallel tempering was used to improve mixing of the high-dimensional
image ensemble. In addition to the target chain,
\[
\Pi_0(\bm{x}) \propto \exp[-\chi^2(\bm{x})/2],
\]
auxiliary chains were evolved at higher temperatures,
\[
\Pi_m(\bm{x})
\propto
\exp[-\chi^2(\bm{x})/(2T_m)],
\qquad
T_0=1 .
\]
The ladder was chosen as \(T_m=T_{m-1}f\), with \(f\simeq1.25\)--\(2\);
about ten replicas were sufficient for the cases studied. Only samples
from the target-temperature chain were used for the reported means,
variances, data-visible standard-deviation maps, and derived observables.

\begin{figure*}[ht!]
\centering
\includegraphics[width=0.95\textwidth]{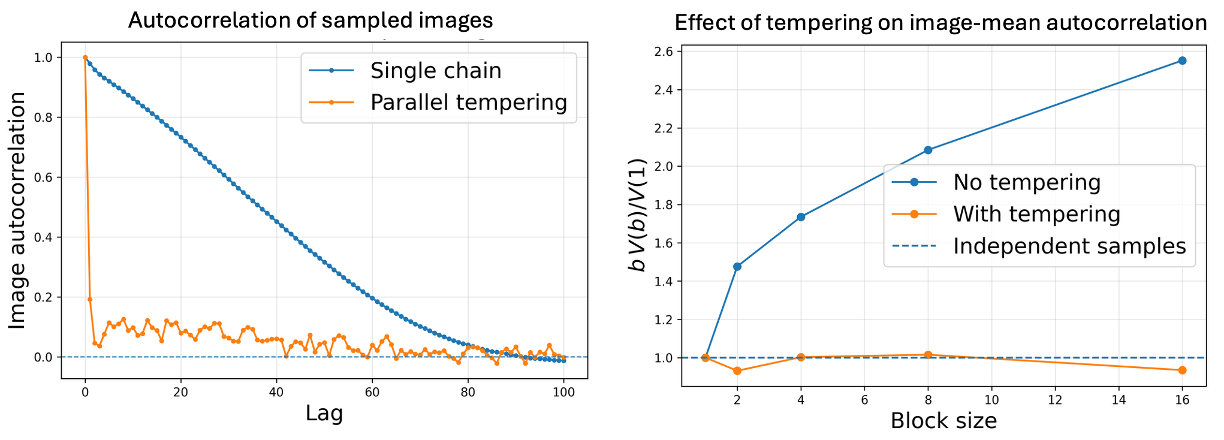}
\caption{
Effect of parallel tempering on HMC mixing. Left: image-level
autocorrelation as a function of lag for a single HMC chain and for the
parallel-tempered run. Right: blocking analysis for an image-mean
observable. For independent samples, \(bV(b)/V(1)\) is expected to remain
close to unity. Parallel tempering substantially reduces serial
correlation and improves effective sampling of the target-temperature
ensemble.
}
\label{fig:hmc_tempering_diagnostics}
\end{figure*}

Mixing was assessed using image-level autocorrelation and blocking
diagnostics. Figure~\ref{fig:hmc_tempering_diagnostics} shows that the
single-chain simulation exhibits a long autocorrelation tail, whereas
parallel tempering rapidly reduces autocorrelation. The blocking analysis
gives the same conclusion: without tempering, \(bV(b)/V(1)\) increases
with block size, indicating serial correlation; with tempering, it remains
close to the independent-sample reference.

% \endgroup
% ===== END SUBSTANTIVE REVISION =====

% \bibliographystyle{iopart-num}
\bibliographystyle{dcu}
\bibliography{bib_PMB_clickable_DOI}

\end{document}